\documentclass{article}

\usepackage{amsfonts,mathtools}
\usepackage{bm}
\usepackage{commath}
\usepackage{ulem}
\usepackage{fullpage}
\usepackage{amsmath}
\usepackage{float}

\usepackage{xcolor} 
\newcommand{\itc}[1]{#1} 
\newcommand{\itcc}[1]{#1}

\newcommand{\revision}[1]{{#1}}

\def\calF{{\cal F}}
\def\calG{{\cal G}}
\def\calA{{\cal A}}
\def\Ft{\tilde{F}}
\def\Gt{\tilde{G}}

\def\ub{\bar{u}}
\def\hb{\bar{h}}
\def\qb{\bar{q}}
\def\vb{\bar{v}}
\def\Gb{\bar{G}}

\def\bE{\mathbb{E}}
\def\brho{\boldsymbol{\rho}}
\def\Cbar{\overline{C}} 

\def\uh{\hat{u}}
\def\ut{\tilde{u}}

\title{Parametrization of subgrid scales in long-term simulations of the shallow-water equations using machine learning and convex limiting}

\author{Md Amran Hossan Mojamder\thanks{\texttt{University of Houston, Dept. of Mathematics, mmojamde@cougarnet.uh.edu, ORCID: 0009-0009-0507-5566}}, 
Zhihang Xu\thanks{\texttt{University of Houston, Dept. of Mathematics, zxu29@central.uh.edu, ORCID: 0009-0007-3802-1017}},
Min Wang\thanks{\texttt{University of Houston, Dept. of Mathematics, mwang55@central.uh.edu, ORCID: 0000-0002-5639-6345}},
Ilya Timofeyev\thanks{\texttt{University of Houston, Dept. of Mathematics, itimofey@cougarnet.uh.edu, ORCID: 0000-0002-3978-4047}}}

\begin{document}

\maketitle

\begin{abstract}
We present a method for parametrizing sub-grid processes in the Shallow Water equations. We define coarse variables and local spatial averages and use a feed-forward neural network to learn sub-grid fluxes. 
Our method results in a local parametrization
that uses a \itc{four-point} computational stencil, which has several advantages over globally coupled parametrizations. 
We demonstrate numerically that our method improves energy balance in long-term turbulent simulations and also accurately reproduces individual solutions. 
\revision{The long-term simulations refer to numerical studies where a fluid flow is simulated over a duration long enough to reach a statistical steady state.}
The neural network parametrization can be easily combined with flux limiting to reduce oscillations near shocks. More importantly, our method provides reliable parametrizations, even in dynamical regimes that are not included in the training data.
\end{abstract}


\smallskip


\section{Introduction}
Simulations of atmosphere-ocean models involve solving numerically complex partial differential equations (PDEs) that span a wide range of temporal and spatial scales. However, due to limited computational resources, climate models may not be able to fully resolve all important physical processes (e.g., \cite{Schneider2017}). \itc{To manage this complexity, coarse models have been developed that use} \textit{parameterizations} 
to approximate the effects of unresolved subgrid-scale processes. These parameterizations can introduce considerable uncertainty and bias into climate predictions (e.g., \cite{WilcoxDonner2007, Bechtold2008, FarnetiGent2011, StevensBony2013, Griffies2015}).

A promising approach is to use Machine Learning to learn improved parameterizations by training on outputs from high-resolution simulations that capture the subgrid dynamics more accurately. 
It has been demonstrated across many applications that \revision{machine learning} models can learn complex relationships by minimizing prediction errors over training datasets.
Among \revision{machine learning} techniques, Neural Networks (NNs) have become particularly popular in climate modeling. They have been applied to various problems such as solving the forced Burgers’ equation~\cite{Alcala2021, Subel2021}, ocean modeling~\cite{BoltonZanna2019, ZannaBolton2020, GuillauminZanna2021}, cloud representation~\cite{Rasp2018}, residual heating and moistening~\cite{BrenowitzBretherton2019}, convection~\cite{Krasnopolsky2013, Yuval2021}.
\revision{Beyond large-scale climate applications, NNs have also shown strong performance 
in hydraulic engineering problems. For example, \cite{Akbar2025Predicting} 
compared several \revision{machine learning} models for predicting scour 
depth in meandering channels and demonstrated that data-driven approaches 
outperform traditional empirical formulas in capturing complex nonlinear 
hydraulic responses.} NNs are advantageous due to their \textit{universal approximation properties}~\cite{Cybenko1989, Hornik1989}, which enable them to represent complex nonlinear mappings with high accuracy.
 \revision{
One representative example 
is the Physics-Informed Neural Network (PINN) framework \cite{raissi2019physics, cuomo2022scientific, cai2021physics, wang2021understanding}, a neural network–based approach that has been widely recognized for its broad applicability to solving PDE problems arising in numerous areas \cite{jin2021nsfnets,hao2026variational, hu2023predicting,thawon2026physics}.}
Moreover, once trained, neural networks enable fast and efficient inference, making them suitable for repeated evaluation within large-scale climate and geophysical simulations. 
Such implementations can further benefit from modern GPU and CPU architectures, particularly when using reduced-precision arithmetic~\cite{Vanhoucke2011}.

Several issues arise when applying neural networks in the context of partial differential equations. 
\revision{For physics-informed models, while they offer flexibility and often reduce the need for large training datasets, 
neural network solvers
may still face certain optimization challenges \cite{basir2022critical,rathore2024challenges}, particularly for stiff, multiscale, or convection-dominated problems \cite{krishnapriyan2021characterizing,xu2025weak,liu2020multi}.
Moreover, for most data-driven neural network approaches, it is typically necessary to perform fully resolved simulations to generate training data. This may limit their applicability over wide parameter ranges, since high-fidelity simulations can be computationally expensive.}
Thus, to make neural networks computationally efficient and practical, it is important to develop \revision{machine learning} approaches that can be easily generalized to parameters outside of the training regime. \itc{For atmosphere/ocean applications, external forcing is an important parameter, and 
parametrizations should respond adequately to a wide range of forcings outside of the training regime.}
\itc{In the context of fluid dynamics, it is important
to develop parametrizations that are local in physical space.} Local parametrizations are computationally efficient and can be trained and modified 
independently for different spatial locations. \itc{In addition, local parametrizations can be analyzed and trained using localized PDE models instead of large-scale models describing the global behavior of the ocean or atmosphere.}
For instance, it is challenging to develop parametrizations for turbulence near a wall (e.g. \cite{germano91dsgs,lilly92dsgs,hoque2024large}) and local \revision{machine learning} parametrizations \itc{can be developed and fine-tuned using a relatively.
inexpensive fully-resolved simulations.}
In addition, local parametrizations typically require a smaller training and testing datasets, and fully resolved simulations can be performed on a smaller computational domain.
\itc{It is also important to develop \revision{machine learning} methods
that produce physically admissible numerical solutions (e.g., positive water height).}
\itc{It is well-understood that standard NNs may not obey physical conservation laws such as mass or energy conservation, and this can lead to non-physical numerical solutions.} \itc{In this paper, this is mitigated by combining the NN parametrization with flux limiters to ensure physically valid numerical solutions of the coarse reduced model.}
\itc{Flux limiters have been widely used in the context of classical numerical methods to alleviate computational problems in hyperbolic conservation laws due to large gradients.}

\itc{The Shallow Water Equations (SWEs) are a system of hyperbolic partial differential equations for the horizontal velocity and the water height. These equations describe the flow of a constant-density fluid layer where the horizontal length scale is much larger than the depth.}
\itc{In this paper, we use Neural Networks to learn nonlinear fluxes 
in the coarse discretization of the SWEs.} 
In particular, we utilize a \itc{four-point} stencil as input to a feed-forward Neural Network and reconstruct non-linear fluxes \itc{in the coarse discretization of the SWEs.} The NN parametrization can be effectively interpreted as a higher-order flux approximation. This approximation is less diffusive compared to the standard Lax-Friedrichs discretization on a coarse mesh.
Moreover, the NN parametrization is local since it requires only \itc{four neighboring} points to reconstruct the flux.
\revision{We also demonstrate that our approach allows the application of the NN parametrization to a range of parameters outside of the training regime. In particular, we
demonstrate that the NN parametrization generalizes well in regimes with increased external forcing up to 40\%. In addition, we show that NN parametrization generalizes to simulations with topography and Manning's friction.}

\revision{The main novelty of our approach is that we use neural networks to model fluxes, and not the solution itself. Thus, we believe that the neural network generalizes better for dynamical regimes not included in the training data, including simulations with topography. In addition, we also compare both spectra in long-term stationary simulations and individual solutions, since analysis of energy spectra is essential for many ocean applications.}
Another advantage of our \revision{machine learning} approach is that it can be easily combined with
traditional numerical methods for hyperbolic conservation laws to ensure the admissibility of numerical solutions.
In particular, we combine our NN parametrization with 
Monolithic Convex Limiting (MCL) strategy (see, e.g., \cite{Kuzmin2020, hajduk2022algebraically})
to ensure that the numerical solution is within \itc{the admissible set.}  
We demonstrate numerically that the MCL strategy does not affect the energy spectra in long-term turbulent simulations. On the other hand, the MCL improved the behavior of individual solutions near shocks.

The rest of the paper is organized as follows.
In section \ref{sec:swe}, we introduce the 1D shallow water equations and their numerical discretization. 
In section \ref{sec:method} we introduce coarse variables and subgrid fluxes (section \ref{sec:coarse_subgrid}), 
discuss network architecture and training (section \ref{sec:net}), 
and introduce the MCL strategy (section \ref{sec:MCL}).
Unlike \cite{Alcala2021}, which relies on conditional Generative Adversarial Networks, we employ a feed-forward neural network to construct the subgrid parametrization, resulting in a simpler training procedure and improved computational efficiency.
We present our numerical results in section \ref{sec:num} and summarize our findings in section \ref{sec:conc}.

\section{Problem formulation}
\label{sec:swe}
\subsection{Shallow water equation}

In this paper, we consider one-dimensional
shallow water equations (SWE) \cite{SaintVenant1871}. These equations are suitable for modeling fluid flows where the horizontal length scale is significantly larger than the vertical fluid depth. These equations can be written using a conservative form
\begin{equation}
\label{swe}
\begin{bmatrix}
h\\q
\end{bmatrix}_{t}+
\begin{bmatrix}
q\\
q^2/h+\frac{1}{2}gh^2
\end{bmatrix}_{x}=
\begin{bmatrix}
0\\ \revision{\rho(x,t)}
\end{bmatrix},
\end{equation}
where $g$ is gravitational acceleration, $h(x,t)$ is the fluid depth, $q = hv$ is the discharge, 
$v(x,t)$ is the horizontal fluid velocity, and $\rho(x,t)$ is a large-scale stochastic forcing. It is also possible to include viscosity in these equations to model the internal friction of the fluid.
The first equation represents the conservation of mass, and the second equation is the conservation of momentum when $\rho =0 $.
These equations are supplemented with periodic boundary conditions
\[
h(0,t) = h(L,t), \quad v(0,t) = v(L,t).
\]
The \textit{initial conditions} at time \( t = 0 \) are given by:
$h(x,0) = h_0(x)$ and $v(x,0) = v_0(x)$.

Equation in \eqref{swe} can be written as a conservation law
\[
u_t + (f(u))_x = \revision{\brho},
\]
where 
\[
u = \begin{bmatrix}
h\\q
\end{bmatrix}, \quad 
f(u) = \begin{bmatrix}
q\\
q^2/h + \frac{1}{2}gh^2
\end{bmatrix}, 
\quad \text{and} \quad
\revision{\brho = \begin{bmatrix}
0\\
\rho(x,t)
\end{bmatrix}.}
\]

The \textit{stochastic forcing} term \revision{\( \rho(x,t) \)}, which models large-scale random forcing at the surface of the fluid (e.g., the wind stress), is given by:
\begin{equation}
\label{eq:rho}
\revision{\rho(x,t)} = A \sum_{k \in K} \left[ \alpha_k(t) \cos\left(2\pi k \frac{x}{L}\right) + \beta_k(t) \sin\left(2\pi k \frac{x}{L}\right) \right],
\end{equation}
where $A$ is the amplitude, $K$ is the set of forced Fourier wavenumbers, and 
$\alpha_k$ and $\beta_k$ are coefficients evolving according to the AR(1)
(autoregressive model of order 1)
process 
\begin{equation}
\label{eq:ab}
\alpha_k(t + \Delta t) = \psi \alpha_k(t) + \sigma \epsilon_{k,1}(t), \quad \beta_k(t + \Delta t) = \psi \beta_k(t) + \sigma \epsilon_{k,2}(t),
\end{equation}
where $0 < \psi = 1 - \gamma\Delta t$ with $\gamma>0$ and $\sigma>0$ are AR(1) parameters, and $\epsilon_{k,i}(t)$, $i=1,2$ are i.i.d. Normal $N(0,\Delta t)$ random variables.
Coefficients $\alpha_k$ and $\beta_k$ are time-correlated Normal $N(0,\sigma^2 / (1 - \psi^2))$ random variables. Equation \eqref{eq:ab} can be viewed as a temporal discretization of the Ornstein-Uhlenbeck process. 
In this paper, we use \( K = \{1,2,3\} \), i.e., the first three wavenumbers are forced.

\subsection{Space-Time discretization}
\label{sec:fvd}
In space,
system~\eqref{swe} is discretized on a 
uniform fine-mesh with $\Delta x = L/N_f$ and
cells 
$C_i = [x_{i-1/2}, x_{i+1/2}]$, $i=0,\ldots,N_f-1$. The 
endpoints of each interval are defined as
$x_{i-1/2 } = i\Delta x$ and $x_{i+1/2} = (i+1) \Delta x$.
The midpoints of each interval $C_i$ are defined as
$x_i = (i + 1/2)\Delta x$. We assume that  $\Delta x$ is small enough and the fine-mesh discretization resolves all physical processes of interest. Thus, we 
use a standard Local Lax-Friedrichs (LLF) (e.g. \cite{leveque2002finite}) space discretization
where the equation \eqref{swe} is discretized as 
\begin{equation}
\label{fvol}
\frac{\dif}{\dif t} u_i = 
-\frac{F_{i+1/2} - F_{i-1/2}}{\Delta x} + \revision{{\brho}_i(t)}, 
\quad i = 0, \ldots, N_f - 1,
\end{equation}
where $u_i \equiv [h_i, q_i]^\top$ is the average over the cell $C_i$.
\revision{
The forcing ${\brho}_i(t) = [0, {\rho}_i(t)]^T$, and we assume that the forcing is slow-varying in space so that $\int_{C_i} \rho(x,t) dx \approx \rho(x_i,t)$.
Thus, we use $\rho_i(t) := \rho(x_i,t)$.}
The LLF flux is vector-valued and is given by 
\begin{equation}
F_{i+1/2}(u_i, u_{i+1}) =\frac{f(u_i) + f(u_{i+1})}{2} - \frac{\lambda_{i+1/2}}{2}(u_{i+1} -u_i) \equiv \calF(u_{i+1}, u_i),
\label{eq:llf}
\end{equation}
where scalar
\begin{equation}
\label{eq:lambda}
\lambda_{i+1/2} = \max\left( |v_i| +\sqrt{gh_i}, \, |v_{i+1}| + \sqrt{gh_{i+1}}\right)
\end{equation}
represents the upper bound for the local wave speed. 
The second term in equation \eqref{eq:llf} represents additional viscosity and ensures the stability of the space discretization. In particular, it is known that the LLF scheme does not produce spurious oscillations near shocks.
The scheme in \eqref{eq:llf} is symmetric, i.e., there is a function $\calF$
\itc{such that right and left 
fluxes can be expressed as $F_{i+1/2} = \calF(u_i,u_{i+1})$ and $F_{i-1/2} = \calF(u_{i-1},u_{i})$.}

The LLF scheme can be rewritten using bar states; this is useful in the context of Monolithic Convex Limiting. The bar states represent spatially averaged exact solution of the Riemann problem for the shallow water equations.
\revision{These states serve as the main building block of the Monolithic Convex Limiting algorithm \cite{Kuzmin2020,hajduk2022bound} discussed in section \ref{sec:MCL}. }
If we define the intermediate bar states as
\begin{equation*}
\overline{u}_{i+1/2} = \frac{u_{i+1} + u_i}{ 2} - \frac{1}{2\lambda_{i+1/2}}(f(u_{i+1}) - f(u_i))\,,
\label{eq:bar_state}
\end{equation*}
then the LLF scheme can be written as
\begin{equation*}
\label{fvol2}
 \frac{\dif}{\dif t} u_i 
 = \frac{1}{\Delta x} [\lambda_{i-1/2} (\overline{u}_{i-1/2} - u_i)+ \lambda_{i+1/2}(\overline{u}_{i+1/2} - u_i)] + \revision{\brho_i(t)}\,,
\end{equation*}
for $i=0,1,\ldots, N_f-1$.
\revision{An important property of the bar states is that they satisfy 
$\overline{u}_{i+1/2} \in \calA$ if $\calA$ is an invariant set of the shallow water equations and $u_i, u_{i+1} \in \calA$.}

In this work, we use Heun's time-stepping method. 
For simplicity, we denote the right-hand side of the semi-discrete system~\eqref{fvol} as \( g(u(x,t)) \), and consider the initial value problem
\begin{equation*}
    u_t = g(u(t)), \quad u(0) = u^0,
\end{equation*}
where \( u(t) \coloneqq [u_1(t), \ldots, u_{N_f}(t)]^\top \). We denote \( u^k = u(t_k) \) as the numerical solution at discrete time \( t_k \), with a fixed time step \( \Delta t = t_{k+1} - t_k \).
Heun's method computes \( u^{k+1} \) from \( u^k \) in two steps:
\begin{align*}
    \text{(Predictor step):} \quad & \tilde{u}^{k+1} = u^k + g(u^k) \Delta t, \\
    \text{(Corrector step):} \quad & u^{k+1} = u^k + \frac{1}{2} \left( g(u^k) + g(\tilde{u}^{k+1}) \right) \Delta t.
\end{align*}
This method provides a second-order accurate time integration while maintaining an explicit scheme. 
Heun's method belongs to the class of strong stability preserving (SSP) methods \cite{shu1988efficient,gottlieb2001strong,gottlieb2011strong}
\revision{with an appropriate condition on the time-step, $\Delta t$. 
In particular, Heun's method
maintains stability when used with spatial discretizations that are stable under forward Euler time-discretization; the \revision{strong stability preserving} coefficient for the Heun's method is $c=1$. Therefore, in most practical applications, the CFL condition is sufficient to guarantee stability. Moreover, if the CFL condition is satisfied, then it is also possible to show that 
$u_i^{k+1} \in \calA$ (set of admissible solutions) since it can be expressed as a convex combination of admissible values $u_i^k, \, \overline{u}_{i-1/2}, \, \overline{u}_{i+1/2} \in \calA$.}
Thus, the resulting space-time scheme is also positivity preserving for the water height.
Time-splitting is used to add the stochastic forcing, \revision{$\brho_i(t)$}, after the second Heun step. 
\revision{The LLF discretization \eqref{eq:llf} is first order in space, and the Heun time-stepping is second-order accurate in time.}

\section{Methodology}
\label{sec:method}
\itc{Performing fully resolved simulations of the SWEs requires a very fine mesh, which can be computationally expensive. This issue becomes particularly important when studying the long-term behavior of climate systems on a large scale.}
\itc{Therefore, it is desirable to obtain a
coarse-mesh discretization of the SWE in many situations, including climate studies.} 
Here, we address this issue in the context of stationary long-term simulations.
In particular, we develop a coarse model that 
improves the turbulent cascade and energy transfer between different Fourier modes. To this end, we use \revision{machine learning} to develop a sub-grid flux parametrization in a model for spatial averages.

\subsection{Coarse mesh and Subgrid fluxes}
\label{sec:coarse_subgrid}
Previously, we considered the discretization of the SWE of a fine-mesh with $\Delta x = L/N_f$.
Next, we define a coarse mesh
\[
\Cbar_I :=
\bigcup_{j=nI}^{n(I+1)-1} C_j=[x_{nI-1/2},x_{n(I+1)-1/2}], \hspace{1cm} I=0,\ldots,N_c-1
\]
with the number of coarse cells $N_c = N_f/n$ and
the corresponding mesh size $\Delta X=n \Delta x$.
We also define coarse variables by spatial averaging
\begin{equation}\label{eq:U}
    U_I(t)=\frac{1}{n} \sum_{j=nI}^{n(I+1)-1} u_j(t), \quad 
I = 0, \ldots, N_c-1, 
\end{equation}
\itc{which implies $U_I \equiv [H_I, Q_I]^\top$
and $Q_I = V_I H_I$ where $H_I$, $Q_I$, and $V_I$ are the water height, discharge, and velocity on a course mesh.}
Averaging equation \eqref{fvol} and taking into account the flux structure \eqref{eq:llf} and the telescoping sum, we obtain equations for $U_I$
\begin{eqnarray}
    \frac{d}{dt}U_I &=& 
- \frac{{F}_{n(I+1)-1/2} - {F}_{nI-1/2}}{\Delta X} + \revision{\brho^{U}_I} \nonumber \\ 
&=&
- \frac{\Ft_{n(I+1)-1/2} - \Ft_{nI-1/2}}{\Delta X}
- \frac{F_{n(I+1)-1/2}^{visc} - F_{nI-1/2}^{visc}}{\Delta X} + \revision{\brho^{U}_I},
\label{eq:reduced1}
\end{eqnarray}
where the fluxes $F_{n(I+1)-\frac12}$ and $F_{nI-\frac12}$ at the two interfaces of the coarse cell $\Cbar_I$ are each decomposed into two parts: a nonlinear part of the fine-mesh flux, denoted by $\widetilde{F}_{n(I+1)-\frac12}$ and $\widetilde{F}_{nI-\frac12}$, and a viscous flux, denoted by $F^{visc}_{n(I+1)-\frac12}$ and $F^{visc}_{nI-\frac12}$, respectively.
\itc{Considering equation \eqref{eq:llf}, the nonlinear part of the flux is given by 
$\Ft_{i-1/2} = (f(u_{i-1}) + f(u_i))/2$
and the viscous flux is given by 
$F_{i-1/2}^{visc} = \lambda_{i-1/2}(u_{i} -u_{i-1})/2$.}
The equation above is exact, but not closed since ${F}_{nI-1/2} = \calF(u_{nI-1}, u_{nI})$ and ${F}_{n(I+1)-1/2} = \calF(u_{n(I+1)-1}, u_{n(I+1)})$ depend on fine-mesh variables. 
\revision{The forcing $\brho^{U}_I := [0,\rho^U_I]^T$ is the average of the stochastic forcing over the coarse cell
$\Cbar_I$, i.e. 
\[
\rho^U_I = \frac{1}{n} \sum_{j=nI}^{n(I+1)-1} \rho_j(t), \quad 
I = 0, \ldots, N_c-1.
\]}
\revision{
In this paper, our goal is to develop closures for the nonlinear and viscous parts of the right-hand side. We assume that the forcing is slowly varying in space, so that the averaged forcing does not deviate significantly from the fine-grid forcing terms in the interval $\Cbar_I$, i.e. 
\[
\rho^U_I \approx \rho(x_j,t) \quad 
\text{for} \quad j=nI,\ldots, n(I+1)-1.
\]
Therefore, we omit the superscript $U$ in the forcing term for the rest of the paper, and we use $\rho_I(t) := \rho(X_I,t)$ with $X_I = (I+1/2)\Delta X$
in coarse simulations.
}

\revision{
Equation \eqref{eq:U} defines the resolved variables in coarse simulations, and sub-grid variables can be defined as fluctuations 
\[
\ut_i = u_i - U_{I(i)}.
\]
Equation \eqref{eq:reduced1} can be rewritten using coarse variables $U_I$
and fluctuations $\ut_i$, and the goal of sub-grid modeling is to find an appropriate closure that eliminates fluctuations $\ut_i$ from the equations
for coarse variables.}

We can rewrite equation \eqref{eq:reduced1} as 
 \begin{equation*}
 \label{reduced2}
    \frac{d}{dt}U_I= - \frac{F_{I+1/2}^c - F_{I-1/2}^c}{\Delta X}+ \brho_I,
\end{equation*}
where the "true" coarse fluxes are given by 
$F_{I+1/2}^c = \Ft_{I+1/2}^c + F_{I+1/2}^{c,visc}$
with 
\begin{equation*}
\label{eq:trueflux}    
\Ft_{I+1/2}^c = \Ft_{n(I+1)-1/2} 
\text{~~and~~} 
F_{I+1/2}^{c,visc} = F_{n(I+1)-1/2}^{visc}, 
\end{equation*}
and similarly for 
$F_{I-1/2}^c$.

The goal of sub-grid modeling is to find functions
$G_{I+1/2}(\vec{U})$ and $G_{I-1/2}(\vec{U})$
(depend only on coarse variables)
that approximate fluxes $F_{I+1/2}^c$ and 
$F_{I-1/2}^c$. 
Here, we treat the nonlinear and viscous parts of the subgrid flux separately and develop different subgrid models. In particular, we use a nonlinear function (neural network) to estimate the nonlinear flux $\Ft_{I+1/2}^c$ and model the viscous flux $F_{I+1/2}^{c,visc}$ using the "stabilizing" LLF term. The resulting sub-grid model is given by
\begin{eqnarray}
    \Ft_{I+1/2}^c &\approx& G_{I+1/2}(\vec{U}) \equiv \calG(U_{I-1}, U_I, U_{I+1}, U_{I+2}),\label{subflux1} \\
    F_{I+1/2}^{c,visc} &\approx& G_{I+1/2}^{visc} \equiv
    \frac{\Lambda_{I+1/2}}{2}(U_{I+1} - U_I)\,,
    \label{subflux2}
\end{eqnarray} 
with
\[
\Lambda_{I+1/2} = \max\left( |V_I| +\sqrt{gH_I}, \, |V_{I+1}| + \sqrt{gH_{I+1}}\right).
\]

With this approximation, the reduced model takes the following form
\begin{equation}
\label{reduced3}
    \frac{d}{dt}U_I= - \frac{G_{I+1/2} - G_{I-1/2}}{\Delta X} - 
    \frac{G_{I+1/2}^{visc} - G_{I-1/2}^{visc}}{\Delta X} + \brho_I.
\end{equation}
In \cite{amranmojamder2026}, we demonstrated that linear regression can be used to justify equation \eqref{subflux2}. At the same time, linear regression indicates that the nonlinear flux cannot be well approximated
in the usualy way, i.e. $\Ft_{I+1/2}^c \not\approx \calF(U_I, U_{I+1})$, especially near shocks.

To develop an accurate approximation of 
$\Ft_{I+1/2}^c$, we utilize a neural network to estimate $\calG$ such that
$G_{I+1/2} = \calG(U_{I-1},U_I,U_{I+1},U_{I+2})$
and $G_{I-1/2} = \calG(U_{I-2},U_{I-1},U_{I},U_{I+1})$. Thus, we develop a symmetric four-point stencil 
approximation for the nonlinear part of the flux $\Ft_{I+1/2}^c$.
A straightforward approach would be to consider a two-point stencil for $\calG$, i.e.
$G_{I+1/2} = \calG(U_{I}, U_{I+1})$.
However, relying solely on a two-point stencil for computing the subgrid fluxes resulted in a suboptimal performance of the coarse model. Thus, we utilized a four-point stencil, as discussed earlier. 
Similar findings were reported in \cite{dta12,dat12,zadoacti18,Alcala2021}, where a two-point stencil was found insufficient for developing an accurate sub-grid approximation.
In particular, an analytical mode-reduction 
approach for stochastic multiscale systems was utilized in \cite{dta12,dat12,zadoacti18} and 
provided a rigorous justification for using the four-point stencil.
This limitation likely arises because using only two points does not provide adequate information for estimating the local curvature of the solution. By increasing the stencil size to include additional neighboring resolved modes, the sub-grid model can better determine whether a sub-grid flux computation is required within a steep gradient or a smoother region.

\subsection{Network Architecture, Dataset, and Training}
\label{sec:net}
To approximate the subgrid flux function \( \mathcal{G} \) mentioned above, we employ a \textit{feedforward neural network} (FNN). 
We consider a set of vector-valued functions \( \widetilde{\calG}_1, \ldots, \widetilde{\calG}_p \) and the target function \( \calG \) can be expressed as the composition 
\[
\calG = \widetilde{\calG}_p \circ \cdots \circ \widetilde{\calG}_1.
\]
Each function \( \widetilde{\calG}_i \) is referred to as the \textit{$i$-th layer} of the network. In this context, \( \widetilde{\calG}_1 \) is the \textit{input layer} and \( \widetilde{\calG}_p \) is the \textit{output layer}, and each function in between is referred to as a \textit{hidden layer}.

\begin{figure}[h]
    \centering
\includegraphics[width=0.85\textwidth]{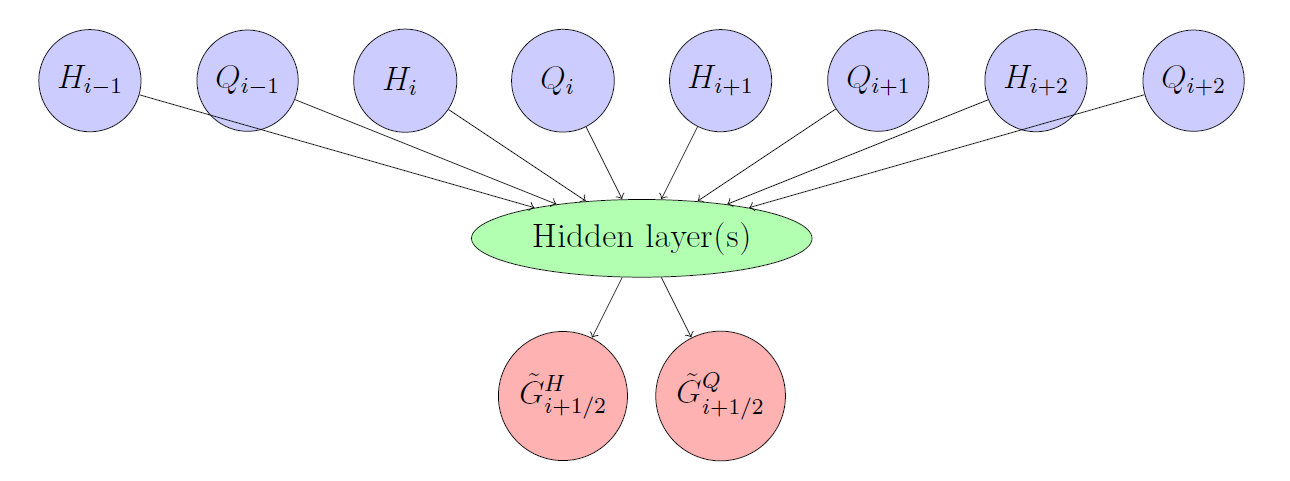}
    \caption{Visual representation of the neural network architecture used to approximate the subgrid flux. The input layer consists of four resolved modes, each with two values (height and discharge). The hidden layers process this input, and the output layer provides the estimated subgrid flux components.}
    \label{fig:net}
\end{figure}

The input layer accepts an 8-dimensional input, and the output layer computes two fluxes (for $H$ and $Q$). Thus, our subgrid approximation is local, and the sub-grid flux approximation can be trained and evaluated independently for different sub-domains, if necessary. We use 3 hidden layers with 128 neurons each.
\revision{To examine the sensitivity of the proposed closure model to architectural choices, we conduct additional experiments by varying the network width, depth, and activation function. For network width, reducing the number of neurons to 64 per hidden layer resulted in a clear degradation in predictive accuracy, indicating insufficient expressive capacity to capture the nonlinear flux corrections. Increasing the width to 256 neurons yielded comparable or slightly improved performance, but with increased computational cost and no substantial long-term stability benefit.
For network depth, using only two hidden layers led to noticeable instability in long-time simulations, suggesting that sufficient depth is required to represent the closure dynamics robustly. Increasing the depth beyond three layers did not produce significant accuracy improvements and introduced additional training complexity.}
We use the GELU (Gaussian Error Linear Unit) 
\cite{hendrycks2016GELU}
activation function to address limitations associated with other activation functions. Unlike ReLU or leaky ReLU, GELU introduces regularization by smoothly blending the linear and nonlinear behavior. The GELU activation is defined as
\[
f(z) = z \cdot \Phi(z),
\]
where \( \Phi(z) \) is the cumulative distribution function (CDF) of the standard normal distribution. This formulation allows the activation to approximate the identity for large positive inputs while smoothly suppressing negative inputs. 
\itcc{GELU has been shown to enhance performance in deep neural networks by providing smooth, differentiable activation with input-dependent gating that improves gradient flow during training (e.g., \cite{Lee2023GELU}).}
\revision{
We further compared activation functions and observed that ReLU-based architectures exhibited reduced stability and limited smoothness in the learned corrections. In contrast, GELU provided smoother nonlinear mappings and improved gradient flow, leading to more stable long-term predictions.
Overall, these experiments indicate that while moderate architectural variations are feasible, the chosen configuration (three hidden layers with 128 neurons and GELU activation) provides a balanced trade-off between expressive capacity, stability, and computational efficiency.}
Figure \ref{fig:net} schematically illustrates the network architecture, with hidden neurons omitted for simplicity.

\textbf{Training Dataset.}
We generate the training dataset by simulating the fine-mesh discretization of the SWE in \eqref{fvol} with $N_f = 1024$ 
and computing the 
coarse variables $U_I$ and
"true" fluxes $\Ft_{I+1/2}^c$.
Since we consider a spatially-homogeneous problem, it is sufficient to compute the training data only for $I=0$.
We use the Heun time-stepping with $\Delta t=0.01$ and simulate each trajectory for the total time $T=400$. We sample data with the time-step $\Delta t^{sample} = 0.2$. Thus, each trajectory generates 2000 snapshots.
It is important to note that we do not include data for the intermediate stage of Heun's method. We generate 100 trajectories with different initial conditions generated randomly as follows
\begin{align}
    & h_0(x)=H_0+A_h \left( \sin \left(\frac{2\pi  x}{L}+\phi_h^{(1)} \right) + \sin\left(\frac{4\pi x}{L}+\phi_h^{(2)}\right) \right) \,, \label{eq:ic1}\\
    & v_0(x)=V_0 \,,\label{eq:ic2}
\end{align}
where $x\in [0,L]$ with $L=100$. The height average $H_0=2$ is constant for all initial conditions. The remaining parameters are randomly generated following the distributions
\begin{equation*}
    V_0=\textit{Unif}[1,2], 
    \hspace{1cm} A_h=\textit{Unif}[0.1,0.6],  \hspace{1cm} 
    \phi_h^{(1)},\phi_h^{(2)}=\textit{Unif}[0,2\pi]
\end{equation*}
where $\textit{Unif}[a,b]$ represents the continuous uniform distribution on $[a,b]$.
We use $n=8$ averaging points.

\textbf{Smoothness Indicator.}
Smoothness Indicators (see, e.g., \cite{zhang2016eno} and references therein) have been used in WENO schemes to measure local smoothness of the solution. The smoothness indicator plays a crucial role in high-resolution numerical schemes, particularly in weighted essentially non-oscillatory methods. It measures the smoothness of a function \( u(x) \) in specific regions of the computational domain, ensuring that the scheme adapts its weights to minimize oscillations near discontinuities while maintaining high-order accuracy in smooth regions.

\begin{figure}[H]
\centering
\includegraphics[width=.75\textwidth]{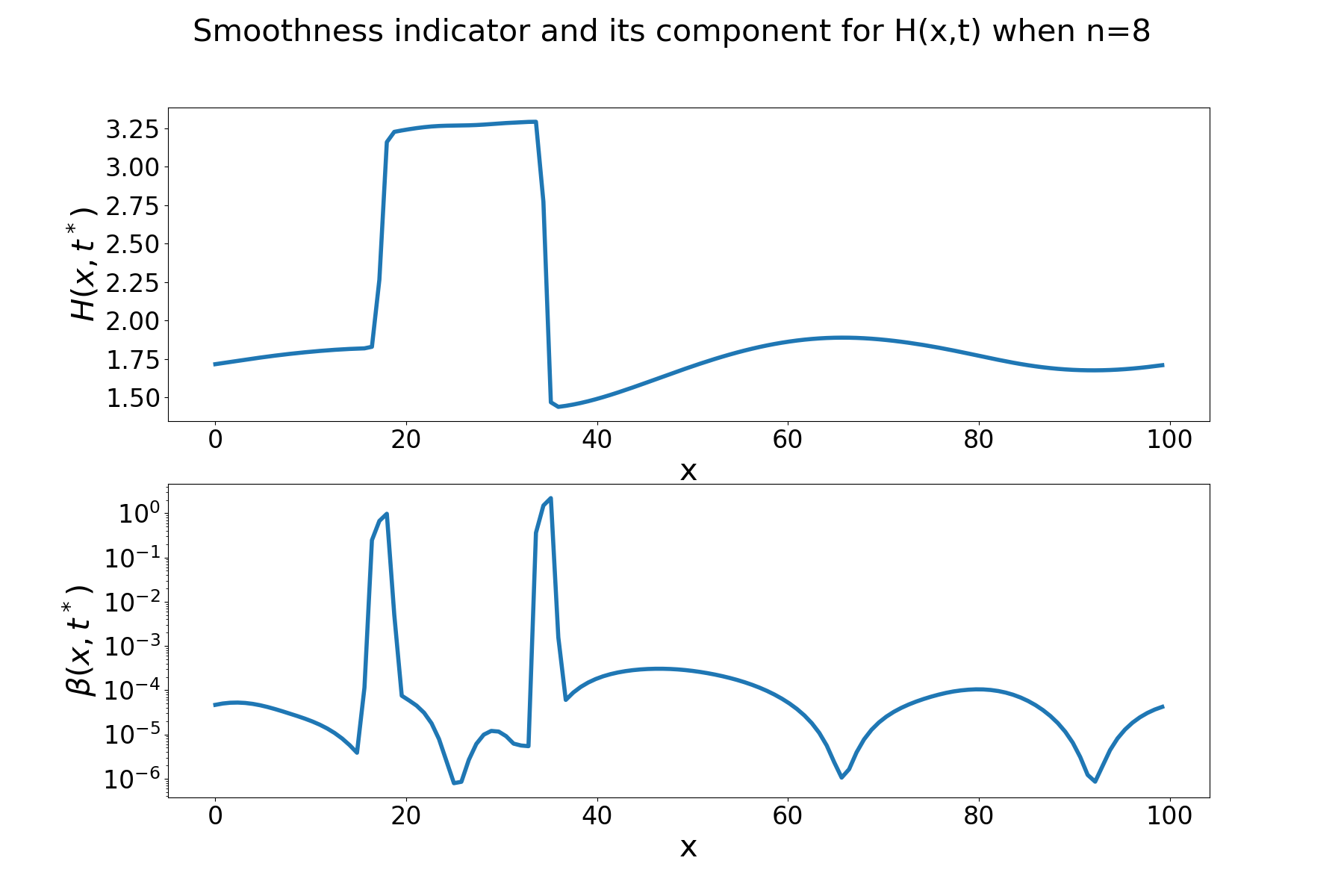}
\caption{Top: a typical profile of the coarse solution $H_I(t)$; Bottom: the corresponding smoothness indicator $\beta_I$ in \eqref{eq:beta}.}
\label{fig:beta}
\end{figure}

The smoothness indicator \(\beta\) is defined as
\begin{equation}
\label{eq:beta}
\beta_i \equiv
\beta(u_{i-1}, u_i, u_{i+1}) = \frac{13}{12} \left( u_{i-1} - 2u_i + u_{i+1} \right)^2 + \frac{1}{4} \left( u_{i} - u_{i+1} \right)^2.
\end{equation}
This formula consists of two terms: the first term, \(\frac{13}{12} \left( u_{i-1} - 2u_i + u_{i+1} \right)^2\), represents the second-order derivative approximation, which captures the curvature of the function. The second term, \(\frac{1}{4} \left( u_{i} - u_{i+1} \right)^2\), represents the square of the first-order derivative, which ensures stability and reduces numerical oscillations. An example of 
the smoothness indicator, $\beta_I$, for a particular snapshot of the coarse solution $H_I(t)$ with $I=0, \ldots, N_c-1$ is presented in Figure \ref{fig:beta}.

\revision{
Figure \ref{fig:beta} demonstrates that the smoothness indicator $\beta$ accurately captures near-shock regions.
The top part of Figure \ref{fig:beta} also indicates that the majority of data  $H_I(t)$, $I = 0,\ldots,N_c-1$ corresponds to smooth regions since there is a relatively small number of shocks and they are highly localized in space.
To improve the robustness of the learning process, we filter the training and testing data based on the smoothness indicator $\beta$. In the full dataset, the distribution of $\beta$ is highly skewed toward small values, corresponding primarily to smooth flow regions. Specifically, the raw dataset has mean $1.35\times10^{-2}$, standard deviation $1.05\times10^{-1}$, and spans a wide range from $8.88\times10^{-12}$ to $7$. Although the maximum value is large due to rare shock events, the 75th percentile is only $4.97\times10^{-4}$, indicating that most of the data corresponds to smooth states. Approximately 25\% of the samples are associated with near-shock regions, identified by $\beta > 4\times10^{-4}$.

It is known that neural networks approximate continuous functions well \cite{Cybenko1989, Hornik1989}, but perform poorly for discontinuous data.
Therefore, training directly on this imbalanced dataset may bias the model toward predominantly smooth states while simultaneously exposing it to extreme shock outliers, which can degrade NN stability. 
To mitigate this issue, we compute the $0.6$ and $0.8$ quantiles of $\beta$ and only retain samples that lie within this interval. This quantile-based filtering concentrates the learning process on moderately active transitional regions, excluding both extremely smooth states and extreme discontinuities. As a result, the neural network is trained on flow regimes where nonlinear effects are significant yet numerically controlled.

The initial unfiltered dataset consists of $3{,}200{,}000$ samples. After applying the $0.6$--$0.8$ quantile filtering, the dataset is reduced to $640{,}000$ samples. The filtered $\beta$ distribution becomes tightly concentrated, with mean $4.02\times10^{-4}$, standard deviation $1.15\times10^{-4}$, and range $[2.40\times10^{-4},\, 6.51\times10^{-4}]$. This substantial reduction in variance confirms that the retained samples correspond to controlled high-gradient regions without extreme shock spikes, thereby improving numerical stability and enhancing predictive robustness.
Finally, we randomly partition the filtered dataset into training and validation subsets: $80\%$ of the samples are used for training, while the remaining $20\%$ are reserved for validation.
}

\textbf{Network Training.}
The training process for the neural network $\tilde{\mathcal{G}}$ follows a structured approach to ensure optimal learning and generalization. Training is performed using the Adam optimizer with a learning rate of 0.001, which facilitates efficient convergence.

\revision{
The loss function employed is the \textit{focal loss}~\cite{lin2017focal}, which encourages the model to focus more on difficult samples (e.g., data near shock regions). 
The classical focal loss was originally proposed for binary classification tasks to address class imbalance. 
Let $y\in\{\pm 1\}$ denote the ground-truth class and $p$ be the model's estimated probability for the class with label $y=1$, 
define \[
p_t = \begin{cases}
    p\,,\quad \text{if } y =1 \,,\\
    1-p\,,\quad \text{otherwise}\,,
\end{cases}
\]
so that $p_t$ represents the predicted probability assigned to the true class. 
The $\alpha$-balanced variant of focal loss is then defined as 
\[
\mathrm{FL}_{p_t} = -\alpha_t(1-p_t)^{\gamma} \log p_t\,,
\]
where $\gamma \ge 0$ is the focusing parameter that 
reduces the impact of well-classified samples and $\alpha_t$ is a class-dependent scaling factor.
For regression tasks, class probabilities are unavailable. Instead, we consider the residual $r_j = \hat{y}_j - y_j$,
where \( \hat{y}_j = G(x_j; \omega, b) \) is the neural network's prediction for the input \( x_j \), and \( y_j \) is the corresponding target value.
We adopt a focal-style weighting mechanism and define the loss function as
\[
C(\omega,b) := \frac{\alpha}{N} \sum_{j=1}^N \left(1 - \mathrm{e}^{-r_j^2}\right)^\gamma r_j^2.
\]
In this formulation, the term $\left(1 - e^{-r_j^2}\right)^\gamma$ acts as an 
adaptive weight. 
For large residuals, the exponential term is negligible, and the loss reduces to a scaled MSE, thereby emphasizing harder examples.
In the present problem, shock regions correspond to rare, spatially localized, 
high-gradient events. The focal weighting reduces the dominance of smooth regions 
during training and encourages improved accuracy in the vicinity of shocks. 
In our experiments, we set $\alpha = 1$ and $\gamma = 2$. 
}

\revision{To mitigate overfitting, early stopping with a patience of 100 epochs was employed, 
with a maximum training budget of 2000 epochs. 
The numerical simulations utilized a single 
compute node equipped with 4 CPU cores and 2 NVIDIA L40S GPUs, each with 46 GB 
of memory. Multi-GPU execution was enabled for training. 
Training was performed using PyTorch, and convergence was typically achieved within the allocated epoch budget. Training of each neural network model required approximately 2–4 hours of wall-clock time.
}

\subsection{Monolithic Convex Limiting}
\label{sec:MCL}
Using a neural network to estimate subgrid fluxes can lead to issues such as under- or overshooting, which may cause negative fluid heights and/or non-physical spurious oscillations near shocks. To prevent this, we apply the \textit{monolithic convex limiting} (MCL) strategy (see, e.g., \cite{Kuzmin2020,hajduk2022bound}). The MCL method allows controlling the subgrid fluxes produced by the neural network, ensuring that the results stay within an admissible set and, in particular, that the fluid height remains positive.
First, we describe the general MCL methodology for the 1D SWE equations, and then we discuss how it is applied to our case.

\subsubsection{General MCL methodology for the 1D SWE.}
\label{sec:MCLgen}
\revision{Here we present the general setup for the MCL methodology. The numerical scheme is presented for the SWE discretized on a general uniform mesh $x_i$.}
Consider a discretization of the SWE of the form
\begin{equation}\label{mcl-flux}
  \frac{d }{dt} u_i = - \frac{(F_{i+1/2} + \Gt_{i + 1/2}^*)
  -(F_{i-1/2}+\Gt_{i-1/2}^*)}{\Delta x},
\end{equation}
where $F_{i \pm 1/2}$ are the LLF fluxes given by equation \eqref{eq:llf} and 
$\Gt_{i \pm 1/2}^*$ are constrained approximations for a higher-order flux correction 
$\Gt_{i \pm 1/2}$. Next, we write \eqref{mcl-flux} in the equivalent bar state form 
\revision{(see \cite{Kuzmin2020,hajduk2022bound} for details)}
\begin{equation}\label{eq:semi-discrete-llfmcl}
  \frac{d}{dt} u_{i}=\frac{1}{\Delta x}\bigl[\lambda_{i-1/2}(\ub_{i-1/2}^{*,+}-u_i)
    +\lambda_{i+1/2}(\ub_{i+1/2}^{*,-}-u_i) \bigr],
\end{equation}
where $\lambda_{i+1/2}$ is the maximum wave speed given by \eqref{eq:lambda}. 
The low-order LLF part 
is given by
\begin{equation*}\label{eq:bar_state-coarse}
\ub_{i+1/2}:=\frac{u_{i+1}+u_i}{2}-\frac{1}{2\lambda_{i+1/2}}(f(u_{i+1})-f(u_i))
\end{equation*}
and
\begin{equation*}
\bar{u}_{i+1/2}^{*,\pm}=  \bar{u}_{i+1/2}\pm\frac{\Gt_{i+1/2}^*}{\lambda_{i+1/2}}.
\end{equation*}

In the SWE context, the admissible set of solutions
consists of all states $u=[h,hv]^\top$ such that $h \ge 0$. 
The bar state $\ub_{i+1/2}$ represents
an averaged exact solution of the Riemann problem with the
initial states $u_i$ and $u_{i+1}$ \cite{guermond2016invariant,hll1983}. 
Therefore, it can be shown that 
$\ub_{i+1/2}$ is in the admissible set if $h_i,h_{i+1} \ge 0$.
\revision{After applying time-stepping scheme, equation \eqref{eq:semi-discrete-llfmcl} indicates that 
$u_i(t+\Delta t)$ can be expressed as a convex combination of 
$\ub_{i-1/2}^{*,+}$, 
$\ub_{i+1/2}^{*,-}$, and
$u_i$. Thus, if these three states are admissible, then $u_i(t+\Delta t)$ is also in the admissible set.}
The MCL approach for the SWE \cite{hajduk2022algebraically} 
ensures that $\ub_{i\pm1/2}^{*,\mp}$ is also admissible
whenever $\ub_{i\pm1/2}$ is admissible 
\revision{with details given in 
\cite{Kuzmin2020,hajduk2022bound}.} 
We will discuss how to construct $\Gt_{i+1/2}^*$ below.

Given that $\ub_{i\pm1/2}^{*,\mp}$
is admissible, each forward stage
of Heun's method advances the initial condition
as follows:
\begin{equation*}
u_i^{\rm MCL} = u_i+\frac{\Delta t}{\Delta x}\bigl[\lambda_{i-1/2}(\ub_{i-1/2}^{*,+}-u_i)
    +\lambda_{i+1/2}(\ub_{i+1/2}^{*,-}-u_i) \bigr].
\end{equation*}
If the time step satisfies the CFL condition
$\Delta t / \Delta x[\lambda_{i+1/2}+\lambda_{i-1/2}]\le 1$,
then $u_i^{\rm MCL}$ is a convex combination of three admissible states. Therefore, for time-steps satisfying the CFL conditions, 
$u_i^{\rm MCL}$ is within the admissible set.

To construct $\Gt_{i+1/2}^*$, 
The MCL algorithm for the SWE system
\cite{hajduk2022algebraically,hajduk2022bound} imposes local bounds on $h$ and $v$. Consider components of the vector-valued flux $
 \Gt_{i+1/2}=[\Gt_{i+1/2}^{h},\Gt_{i+1/2}^{q}]^\top$ and the vector-valued solution
$\ub_{i+1/2}=[\hb_{i+1/2},\hb_{i+1/2}\vb_{i+1/2}]^\top$,
 where $\vb_{i+1/2}:={\qb_{i+1/2}}/{\hb_{i+1/2}}$.
The local bounds are defined as
\begin{equation*}
\begin{aligned}
   & h_i^{\min}:=\min(\bar h_{i-1/2},\bar h_{i+1/2}),
   \qquad h_i^{\max}:=\max(\bar h_{i-1/2},\bar h_{i+1/2}),\\
    & v_i^{\min}:=\min(\bar v_{i-1/2},\bar v_{i+1/2}),
   \qquad v_i^{\max}:=\max(\bar v_{i-1/2},\bar v_{i+1/2})
\end{aligned}
\end{equation*}
and the local maximum principle can be formulated \cite{hajduk2022algebraically} as 
\begin{equation}
\label{mcl-seq}
\begin{aligned}
   h_i^{\min}\le \bar h_{i+1/2}^{*,-}\le h_i^{\max},\qquad
   & h_{i+1}^{\min}\le \bar h_{i+1/2}^{*,+}\le h_{i+1}^{\max},\\
   \bar h_{i+1/2}^{*,-}
   v_i^{\min}\le \bar q_{i+1/2}^{*,-}\le
\bar h_{i+1/2}^{*,-}
v_i^{\max},\qquad
    & \bar h_{i+1/2}^{*,+}
   v_{i+1}^{\min}\le \bar q_{i+1/2}^{*,+}\le \bar h_{i+1/2}^{*,+}v_{i+1}^{\max}.
\end{aligned}
\end{equation}
Since $\hb_{i+1/2}^{*,\pm}=\hb_{i+1/2}\pm
{\Gt_{i+1/2}^{h,*}}/{\lambda_{i+1/2}}$, the water height
constraints can be rewritten as
\begin{equation*}
\begin{aligned}
& \lambda_{i+1/2}(\bar h_{i+1/2}-h_i^{\max})\le 
\tilde G_{i+1/2}^{h,*}\le\lambda_{i+1/2}(\bar h_{i+1/2}-h_i^{\min}),\\
& \lambda_{i+1/2}(h_{i+1}^{\min}-\bar h_{i+1/2})\le
\tilde G_{i+1/2}^{h,*}\le
\lambda_{i+1/2}(h_{i+1}^{\max}-\bar h_{i+1/2}).
\end{aligned}
\end{equation*}
Therefore, it follows that the bound-preserving approximation to
$\tilde G_{i+1/2}^{h}$ is 
\begin{equation}
\label{mcl-limiter-h}
  \tilde G_{i+1/2}^{h,*}=\begin{cases}
  \min(  \tilde G_{i+1/2}^{h},\lambda_{i+1/2}\min(
  \bar h_{i+1/2}-h_i^{\min},h_{i+1}^{\max}-\bar h_{i+1/2}))
  & \mbox{if}\  \tilde G_{i+1/2}^{h}\ge 0,\\
  \max(  \tilde G_{i+1/2}^{h},\lambda_{i+1/2}\max(
  \bar h_{i+1/2}-h_i^{\max},h_{i+1}^{\min}-\bar h_{i+1/2}))
  & \mbox{otherwise}.
  \end{cases}
\end{equation}  
The constraint for the discharge is defined using
$\Delta\Gt_{i+1/2}^{q,*}= \Gt_{i+1/2}^{q,*}-\Gt_{i+1/2}^{h,*} \vb_{i+1/2}$.
The limited flux for the discharge $\Gt_{i+1/2}^{q,*}$ is computed from the limited difference $\Delta\Gt_{i+1/2}^{q,*}$
given by (see \cite{hajduk2022algebraically}) 
\begin{equation*}
\begin{aligned}
\lambda_{i+1/2}\hb_{i+1/2}^{*,-}(\vb_{i+1/2}-v_i^{\max})\le 
\Delta \Gt_{i+1/2}^{q,*}\le\lambda_{i+1/2}\hb_{i+1/2}^{*,-}
(\vb_{i+1/2}-v_i^{\min}),\\
\lambda_{i+1/2}\hb_{i+1/2}^{*,+}(v_{i+1}^{\min}-\vb_{i+1/2})\le
\Delta\Gt_{i+1/2}^{q,*}\le
\lambda_{i+1/2}\hb_{i+1/2}^{*,+}(v_{i+1}^{\max}-\vb_{i+1/2})
\end{aligned}
\end{equation*}
Conditions above can be rewritten as
\begin{equation}
\label{mcl-limiter-q}
  \Delta \tilde G_{i+1/2}^{q,*}=\begin{cases}
  \min(  \Delta\tilde G_{i+1/2}^{q},\Lambda_{i+1/2}\min(
  \bar h_{i+1/2}^{*,-}(\bar v_{i+1/2}-v_i^{\min}),
  \bar h_{i+1/2}^{*,+}(v_{i+1}^{\max}-\bar v_{i+1/2})))
  & \text{if} \,  \Delta \tilde G_{i+1/2}^{q}\ge 0,\\
  \max(\Delta  \tilde G_{i+1/2}^{q},\Lambda_{i+1/2}\max(
  \bar h_{i+1/2}^{*,-}(\bar v_{i+1/2}-v_i^{\max}),
  \bar h_{i+1/2}^{*,+}(v_{i+1}^{\min}-\bar v_{i+1/2})))
  & \mbox{otherwise}.
  \end{cases}
\end{equation}  
Conditions \eqref{mcl-seq} guarantee
positivity preservation and admissibility
of numerical solutions to the 1D SWE.


\subsubsection{Application of MCL in the context of Machine Learning.}
\revision{Here we discuss the application of the MCL algorithm on the coarse grid. Here, the coarse index $I$ should be substituted into all MCL formulas presented in Section \ref{sec:MCLgen}.}
To apply the MCL formalism to our \revision{machine learning} model, we rewrite the model in \eqref{fvol}
as
\begin{equation}
\label{reduced4}
    \frac{d}{dt}U_I= - \frac{\Gb_{I+1/2} - \Gb_{I-1/2}}{\Delta X} -
    \frac{\Gt_{I+1/2}^* - \Gt_{I-1/2}^*}{\Delta X}
    - \frac{G_{I+1/2}^{visc} - G_{I-1/2}^{visc}}{\Delta X} + \brho_I,
\end{equation}
where $\Gb_{I+1/2}$ is the nonlinear part of the LLF flux, and $\Gt_{I+1/2}^*$ is the limited difference between the neural network and $\Gb_{I+1/2}$, i.e.
\[
\Gb_{I+1/2} = \frac{f(U_{I+1}) + f(U_I)}{2}, \qquad 
\Gt_{I+1/2}^* = MCL(G_{I+1/2} - \Gb_{I+1/2}),
\]
with $G_{I+1/2}$ is the Neural Network model given by \eqref{subflux1}.
In the context of MCL formalism in \eqref{mcl-flux} and the reduced model in \eqref{reduced3}, $u_I \equiv U_I$, $F_{I+1/2} = \Gb_{I+1/2} + G_{I+1/2}^{visc}$, and $\Gt_{I+1/2}^*$ 
\revision{is computed using \eqref{mcl-limiter-h} and \eqref{mcl-limiter-q}.} In this context, $\Gt_{I+1/2}^*$ can be interpreted as the limited difference between a low-order and high-order neural network approximations of nonlinear terms.
Therefore, the application of the MCL formalism is rather straightforward - first, the bar states are computed using the LLF scheme, and then the limiter is applied to the difference between the fluxes generated by the neural network and the nonlinear part of the LLF flux.
Time-splitting is used to add the stochastic forcing after the second Heun step.

\section{Numerical results}
\label{sec:num}
In this section, we describe our numerical results for the reduced model with and without the MCL. By analogy with turbulence modeling, we refer to the simulations of the full model as DNS. \itcc{We verified numerically that simulations of the full model with $N_f = 1024$ are adequate for reproducing the solutions of the SWE.} In particular, the numerical diffusion is small and has a minimal effect. Therefore, we use $N_f=1024$ to generate training and testing datasets, and benchmark the performance of our reduced models. 

To train our neural network model, we consider the averaging window $n=8$, which results in the number of coarse degrees of freedom $N_c=128$.  \itc{After training, we also use the same neural network}
\itc{(without re-training) to estimate coarse fluxes} in simulations with 
$N_c=64$, $256$, and $512$ coarse variables. 
This allows us to test how well \itc{our \revision{machine learning} model generalizes with respect to changes in the resolution of coarse equations.}
To distinguish between reduced models with different spatial resolution, the number of coarse degrees of freedom is appended to the abbreviation of the corresponding reduced model.
For instance,
NN reduced models \eqref{reduced3}
\emph{without the limiter} are referred to as NN-64, NN-128, NN-256, and NN-512.
NN reduced models \eqref{reduced4} \emph{with the limiter} are 
referred to as NN-MCL-128 and NN-MCL-256.
We also compare our reduced models with simulations of the straightforward LLF discretization of the SWE in \eqref{fvol} and denote those as LLF-64, LLF-128, LLF-256, and LLF-512.

The length of the domain is \revision{$L=100 m$, the final time is $T=400 s$, and the gravitational acceleration is $g=9.812 m/s^2$.}
We use periodic boundary conditions, and initial conditions are generated according to \eqref{eq:ic1}, \eqref{eq:ic2}.
The Heun method is used to perform the time-stepping for the deterministic part, and Euler time-stepping is utilized to add the stochastic forcing after the second Heun step. The computational time step is $\Delta t = 0.01$. We performed a preliminary investigation and determined that this time step satisfied the CFL condition in all simulations, i.e., 
$\Delta t < \max_{i,t}(\lambda_{i+1/2}) \Delta x$
and $\Delta t < \max_{I,t}(\Lambda_{I+1/2}) \Delta X$.
For the forcing in \eqref{eq:rho}, \eqref{eq:ab} we use $A=0.1$, $\psi=1-\Delta t$, and $\sigma=1.41\sqrt{\Delta t}$, so  that $\alpha_k(t), \beta_k(t) \sim N(0,2/(2 - \Delta t))$ with decaying covariance $\bE[\alpha_k(t + m\Delta t) \alpha_k(t)] \sim \psi^m$. We use the same computational time-step $\Delta t=0.01$ to update $\alpha_k(t)$ and $\beta_k(t)$ after the time-stepping for the SWE.

We generate long stationary time series of dependent variables and compute energy spectra in Fourier space. For any periodic function $u(x,t)$, the Fourier expansion is given by 
$u(x,t) = \sum_k \uh(t) e^{i2\pi kx/L}$,
and the energy spectra is defined as
\begin{equation*}
\label{eq:ek}
e_k = \frac{1}{T} \int\limits_0^T |\uh_k(t)|^2 \, dt.
\end{equation*}
We use $T=400$ and sample data every $\Delta t^{sample} = 0.2$.

\begin{figure}[H]
\centerline{
\includegraphics[width=0.5\textwidth]{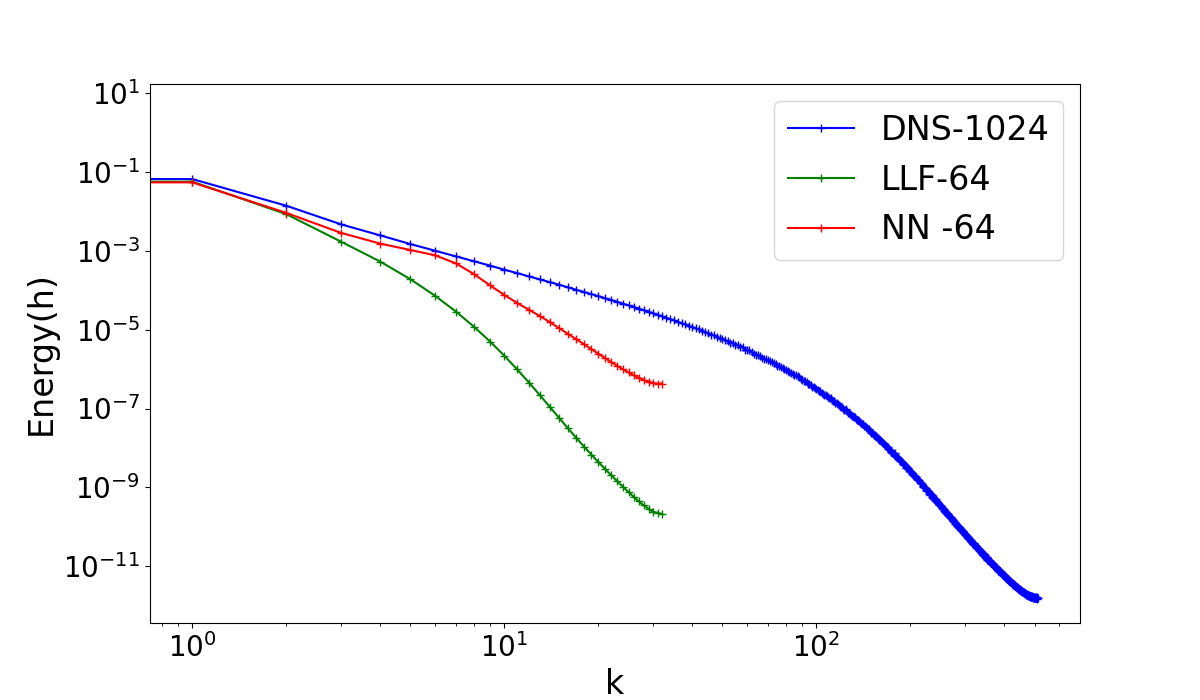}    \includegraphics[width=0.5\textwidth]{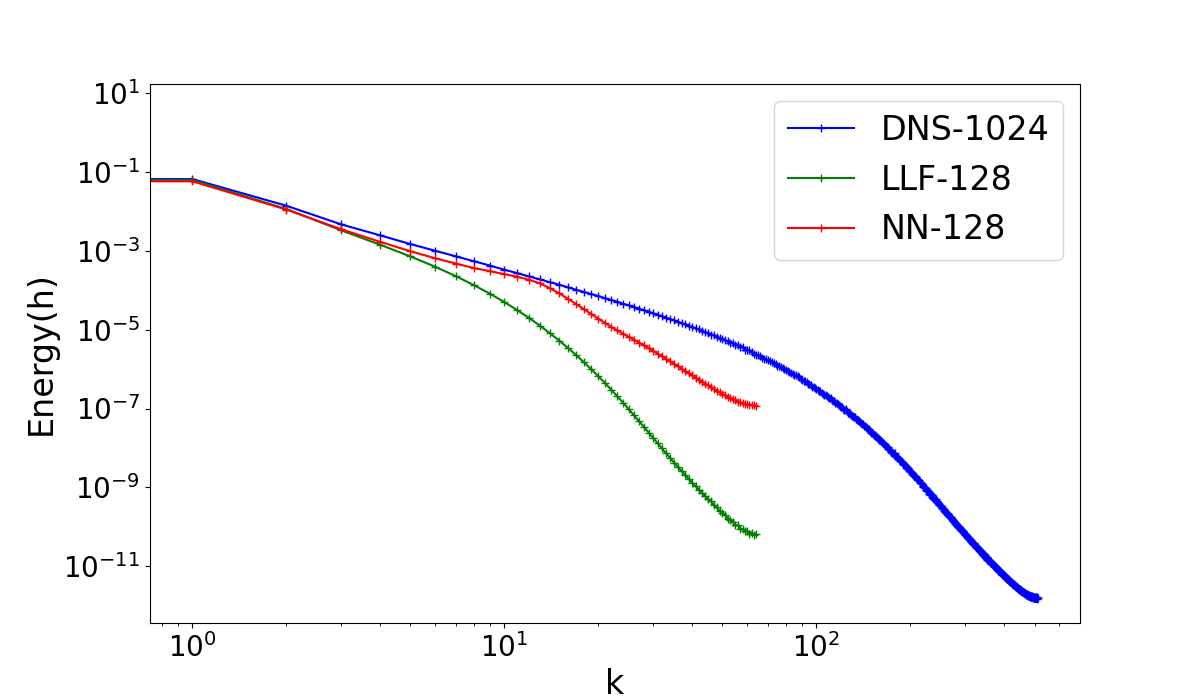}}
\centerline{
\includegraphics[width=0.5\textwidth]{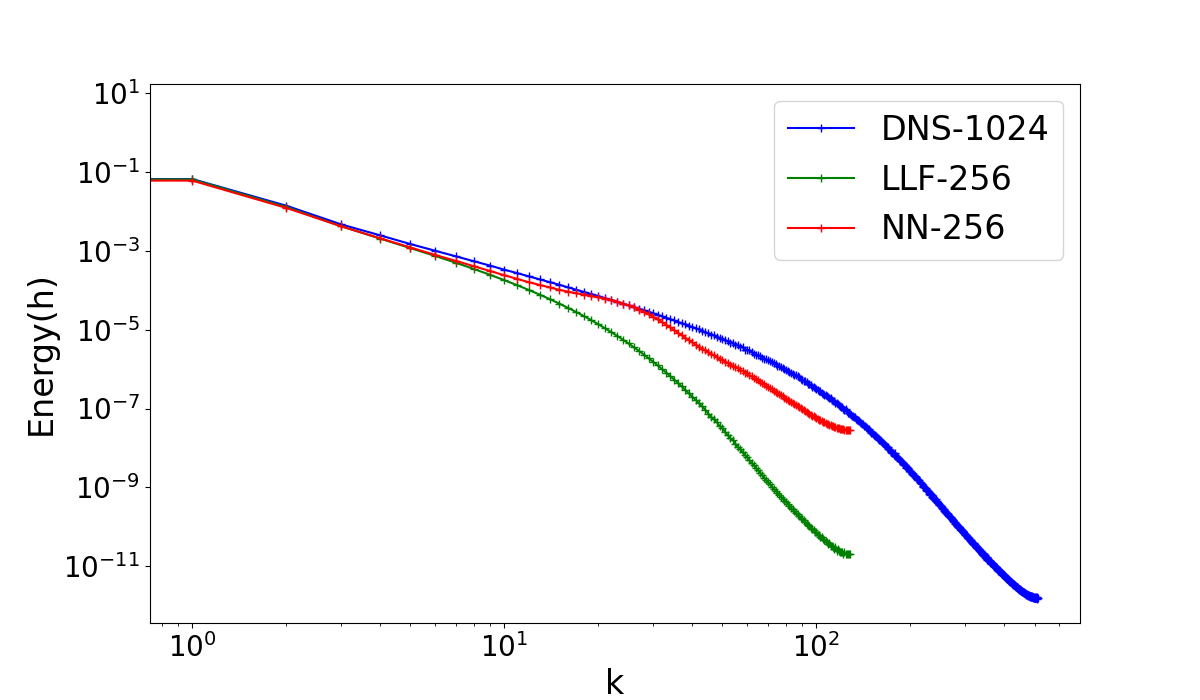}    \includegraphics[width=0.5\textwidth]{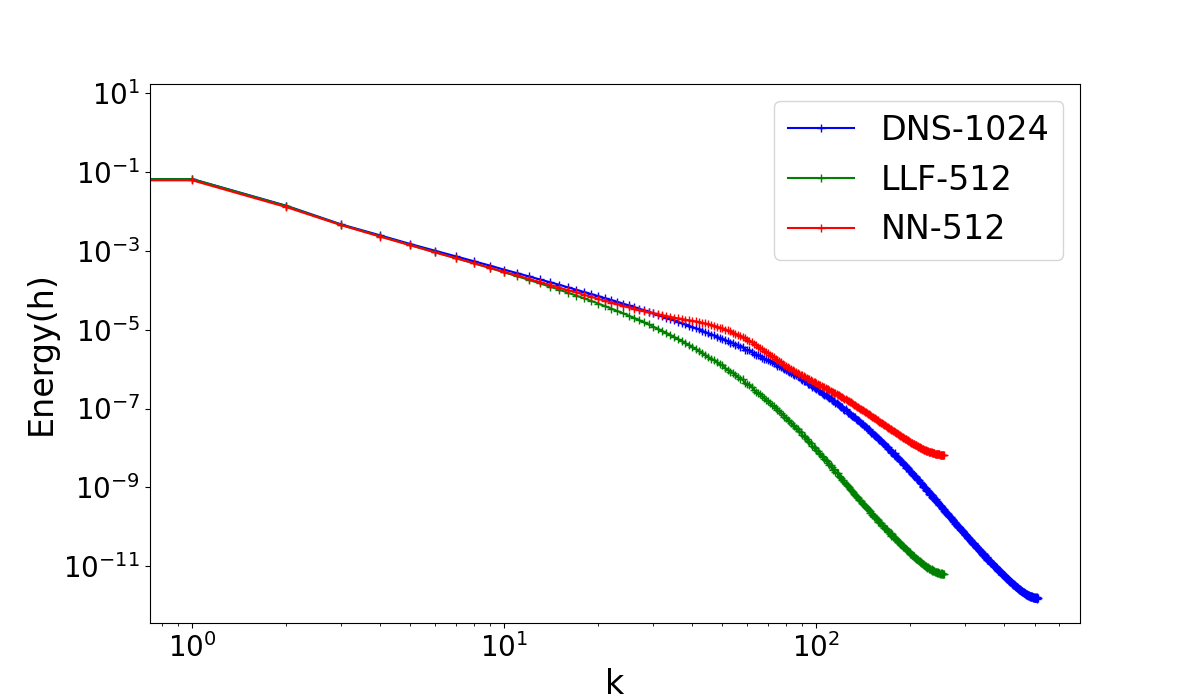}}
\caption{Energy spectra for the water height $h(x,t)$ in long stationary simulations of the full model and reduced models with different resolutions. Note that the reduced \revision{neural network} model is trained in the $N_c=128$ regime and is used for other resolutions without retraining. Blue line - DNS-1024 \revision{(direct numerical simulation with 1024 grid points)}, Red Line - Reduced neural network model with resolution $N_c$, Green Line - \revision{local Lax–Friedrichs solver} with resolution $N_c$. Simulations with $N_c = 64$, $128$, $256$, $512$ - top left, top right, bottom left, bottom right, respectively.}
\label{fig:energy_spectra}
\end{figure}

Figure \ref{fig:energy_spectra} depicts the energy spectra of the water height $h(x,t)$ in the simulations of the full model and reduced models with resolutions $N_c=64$, $128$, $256$, and $512$. \itc{Note that the NN is trained in the regime $n=8$
($N_c=128$).}
At the lowest resolution $N_c=64$, the NN reduced model exhibits the largest discrepancies compared with the fully resolved model. As the resolution increases ($N_c = 128$, $256$, $512$), the NN reduced model aligns more closely with the true spectrum. 
Figure \ref{fig:energy_spectra}
demonstrates that the reduced \revision{machine learning} models performs much better than the LLF discretization with the same spatial resolution. In particular, the NN reduced model reproduces the inertial range of the full model up to wavenumber $k \approx 13$ and $k \approx 37$ for $N_c=128$ and $N_c=256$, respectively. This demonstrates that the NN reduced model improves the representation of energy transfer across scales. However, the NN-128 model is more diffusive compared to the NN-256 model.
Representation of the inertial range in simulations with other values of $N_c$ is also improved by the NN reduced model. Energy spectra for the discharge, $q$, follow a similar trend; some of the results for $q$ are presented later in this paper.
These results indicate that the NN reduced model results in a better approximation of fine-scale structures, ensuring a more accurate representation of energy transfer across scales.
Thus, the neural network considerably improves the missing subgrid dynamics in coarse simulations.

\revision{The NN reduced model conserves the total mass $\int h(x,t) \, dx$ up to a very high precision, less than 0.1\% relative error. This is due to the 
construction of the reduced model in flux form. The kinetic  and potential energies 
(given by $\tfrac12\int uq \, dx$ and $\tfrac12 g \int h^2 \, dx$, respectively)
fluctuate over time in response to the random forcing $\rho(x,t)$ applied to the equation for the discharge. 
The power spectra indicate that the reduced model is slightly more diffusive compared with the fully resolved model, resulting in slightly lower total energy. Indeed, both potential and kinetic energies are slightly smaller in the simulations of the reduced model compared with the fully resolved model if both models are simulated with the identical forcing; this is especially visible at the energy peaks (in time). Largest (peak) relative errors for the potential and kinetic energies are approximately 3-5\% and 10-12\%, respectively \cite{amranmojamder2026}. The MCL does not significantly affect the mass conservation and energy balance.}

\subsection{Simulations with Larger Forcing.} 
Next, we investigate how well the NN reduced model performs outside the training regime. To this end, we consider larger forcing magnitudes of $A=0.12$ and $0.14$, corresponding to 20\% and 40\% increases in the forcing magnitude.  Since forcing is applied to the discharge, $q$, only, an increase in forcing magnitude has a stronger effect on $q$ than on $h$.
Table \ref{tab:energy} summarizes the total spectral energy
$E = \tfrac12 \sum_k e_k$ for $h$ and $q$. As expected, the increase in the energy for $h$ is not significant, but energy for $q$ increase by 27.5\% and 54.6\%, respectively 
due to the nonlinear nature of the equation for the discharge.
The NN reduced model responds correctly to the increase in forcing and, overall, accurately reproduces the energy increase in $q$. The discrepancy between the total energy in the DNS and the NN reduced model is larger for 
$q$ than for $h$, most likely due to the nonlinear fluxes in the equation for the discharge. Similar to the approach in \cite{zadoacti18}, to fix the overall energy budget, it is possible to introduce a single (possibly trainable) parameter to multiply the NN parametrization in the equation for the discharge, 
\itc{which is a straightforward modification of the NN parametrization approach presented here, and we do not discuss it in the present paper.}
\begin{table}[H]
\centering
\begin{tabular}{|c|c|c|c|c|c|c|c|c|c|c|c|c|}
\hline
\multicolumn{1}{|c|}{$N_c$} &
\multicolumn{4}{c|}{$A=0.1$} &
\multicolumn{4}{c|}{$A=0.12$} &
\multicolumn{4}{c|}{$A=0.14$} \\
\hline
& $E_{DNS}^h$ & $E_{NN}^h$  & $E_{DNS}^q$ & $E_{NN}^q$ & 
$E_{DNS}^h$ & $E_{NN}^h$  & $E_{DNS}^q$ & $E_{NN}^q$ &  
$E_{DNS}^h$ & $E_{NN}^h$  & $E_{DNS}^q$ & $E_{NN}^q$ \\
\hline
128 & 
4.093 & 4.078 &  1.858 & 1.564  & 
4.12 & 4.10 & 2.37 & 2.02 &
4.148 & 4.128 & 2.873 & 2.496 \\
\hline
256 & 
4.093 & 4.083 & 1.858 &  1.677 & 
4.12 & 4.11 & 2.37 & 2.17  &
4.148 & 4.136 & 2.873 & 2.655 \\
\hline
\end{tabular}
\caption{Comparison of total spectral energy for $h$ and $q$ for different resolutions, $N_c=128$ and $N_c=256$ in the full model ($E_{DNS}$) and NN reduced model ($E_{NN}$). \revision{The spectral energy is defined as the sum of spectra $E = \tfrac{1}{2}\sum_k e_k$ for $h$ and $q$, respectively. The initial conditions for the water depth are initialized with $H_0=2$. Therefore, the spectral energy for $h$ includes the mean energy; the fluctuating energy can be computed as $E^h - H_0^2$.}
}
\label{tab:energy}
\end{table}

%
%
\begin{figure}[H]
\centerline{
\includegraphics[width=0.48\textwidth]{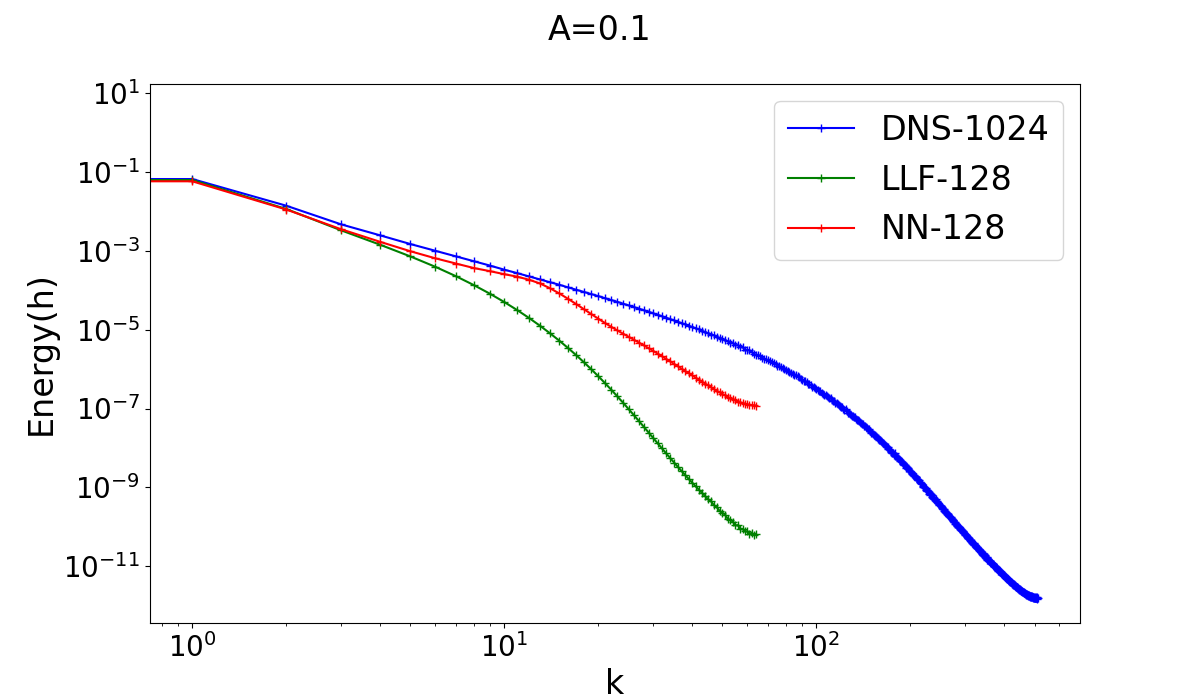}
\includegraphics[width=0.48\textwidth]{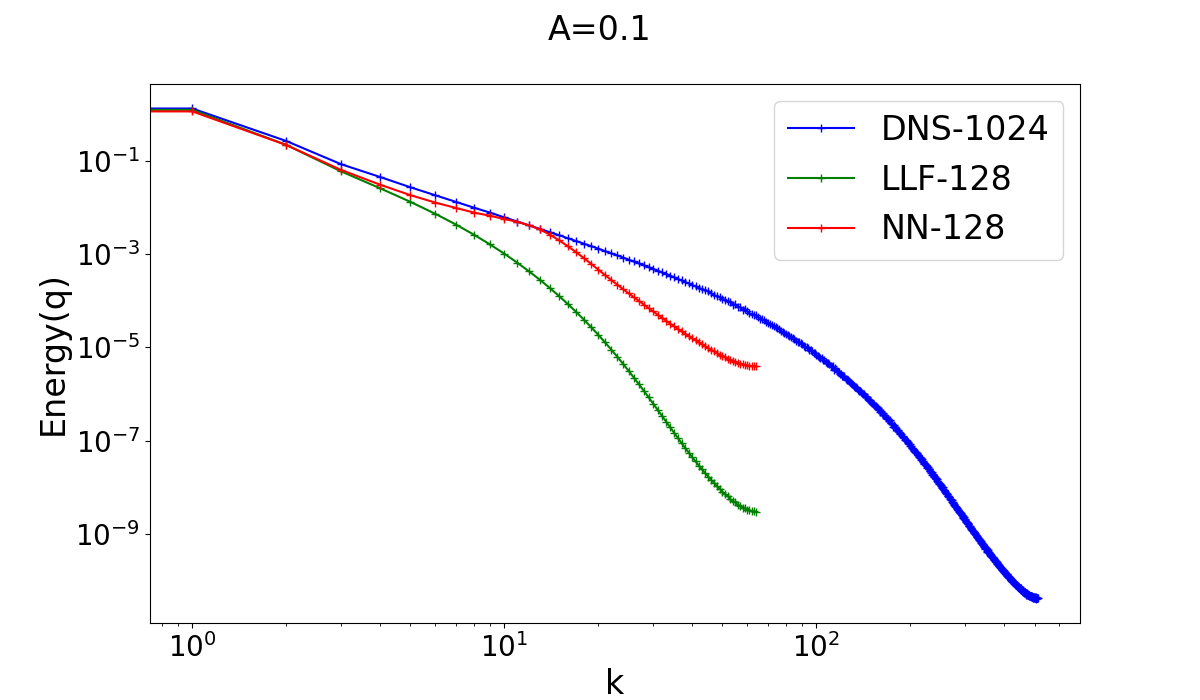}
}
\centerline{
\includegraphics[width=0.48\textwidth]{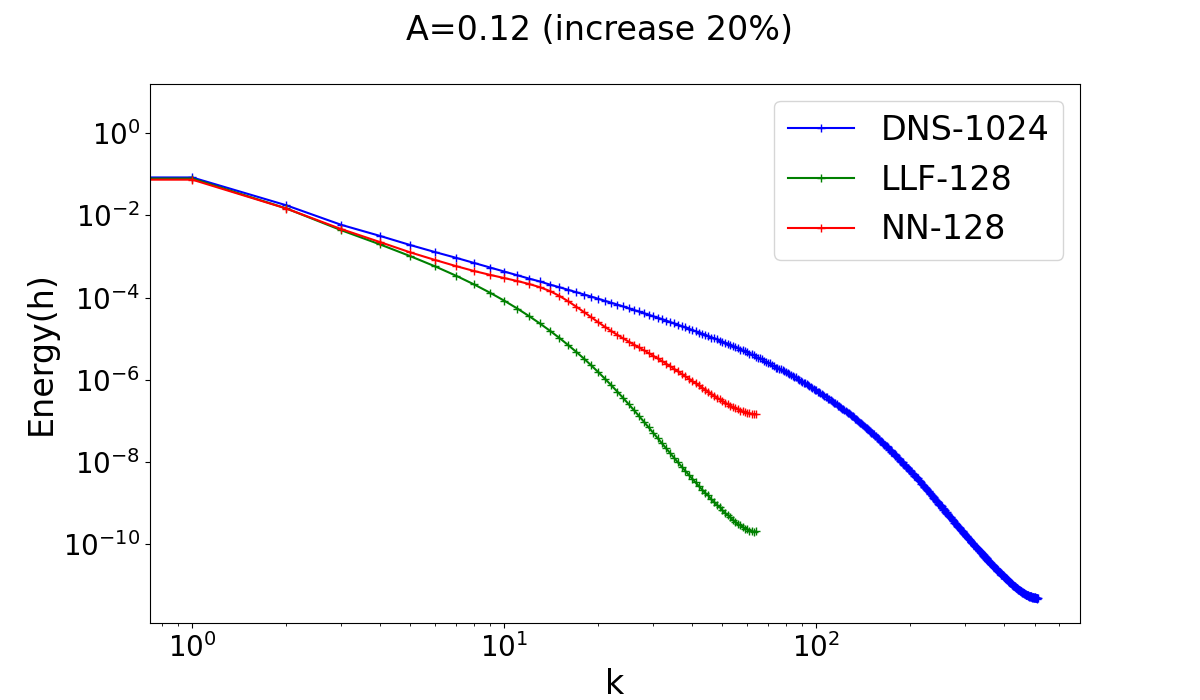}
\includegraphics[width=0.48\textwidth]{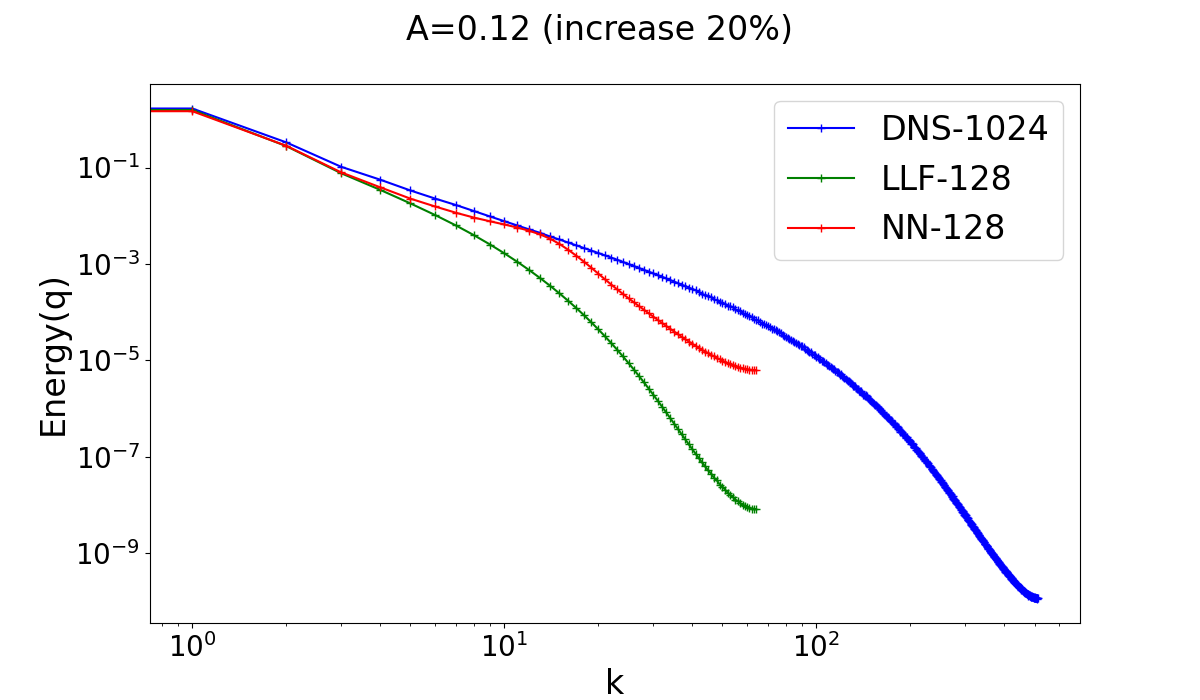}
}
\centerline{
\includegraphics[width=0.48\textwidth]{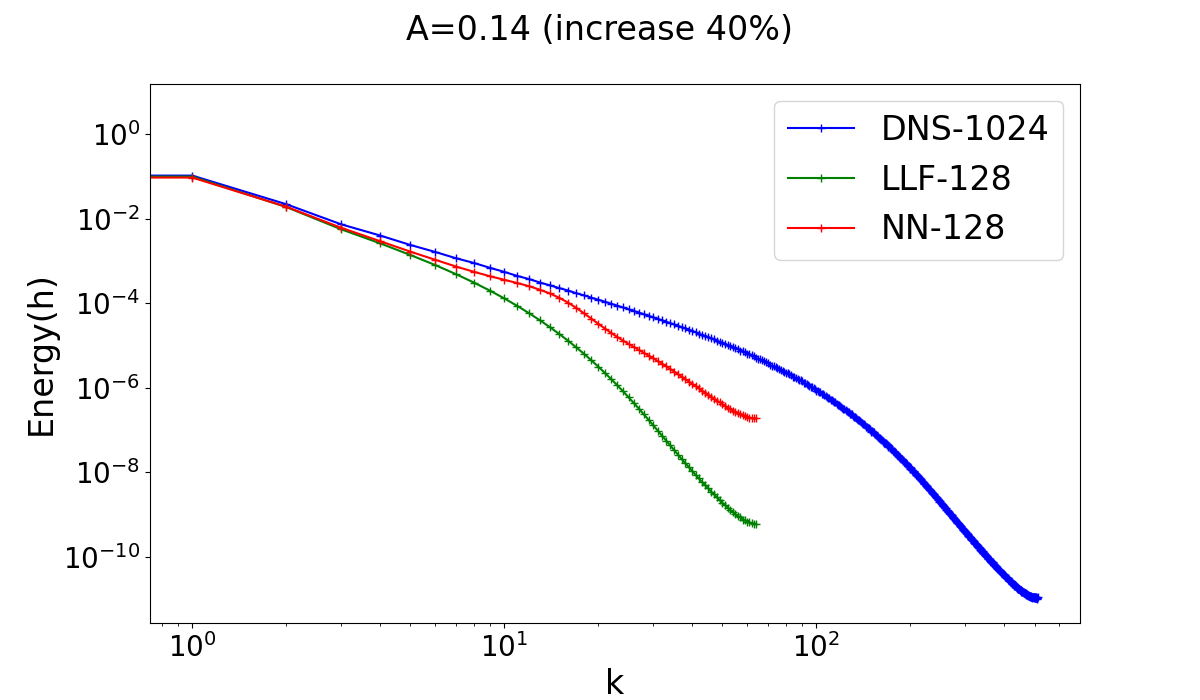}
\includegraphics[width=0.48\textwidth]{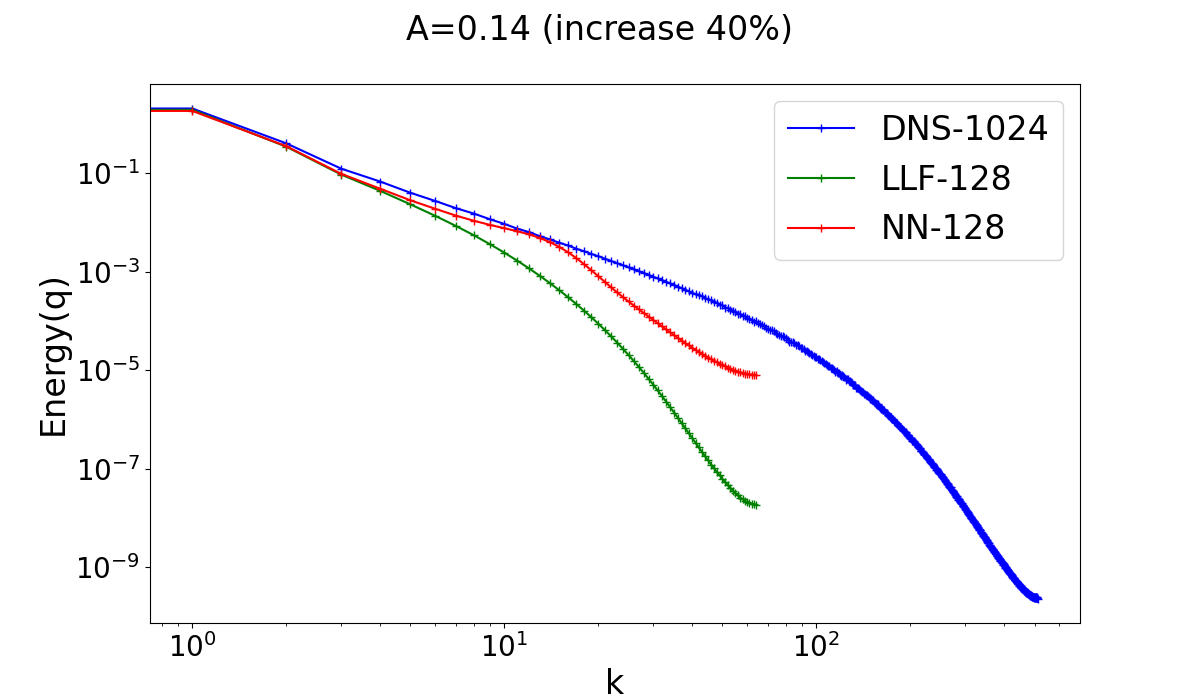}
}
\caption{Energy spectra for $h$ (left) and $q$ (right) in simulations with $N_c=128$ and forcing amplitudes $A=0.1$ (top row), $0.12$ (middle row), and $0.14$ (bottom row). Note that the \revision{neural network} parametrization is originally trained on $A=0.1$ and $N_c=128$; it is used without retraining in other regimes.}
\label{fig:spectra_amplitude}
\end{figure}
Figure \ref{fig:spectra_amplitude} depicts energy spectra for $h$ and $q$ in simulations with $N_c=128$ and $A=0.1$, $0.12$, and $0.14$.
Energy spectra computed from the NN reduced model represent a considerable improvement over coarse LLF discretization, indicating that the neural network effectively reconstructs the subgrid contributions. 
The NN reduced model maintains fidelity for a wide range of forcing amplitudes, demonstrating robustness in capturing the small-scale energy dynamics. 
These results suggest that the NN model can generalize well across varying amplitudes and recover fine-scale structures without retraining.

\begin{figure}[H]
\centerline{
\includegraphics[width=0.45\textwidth]{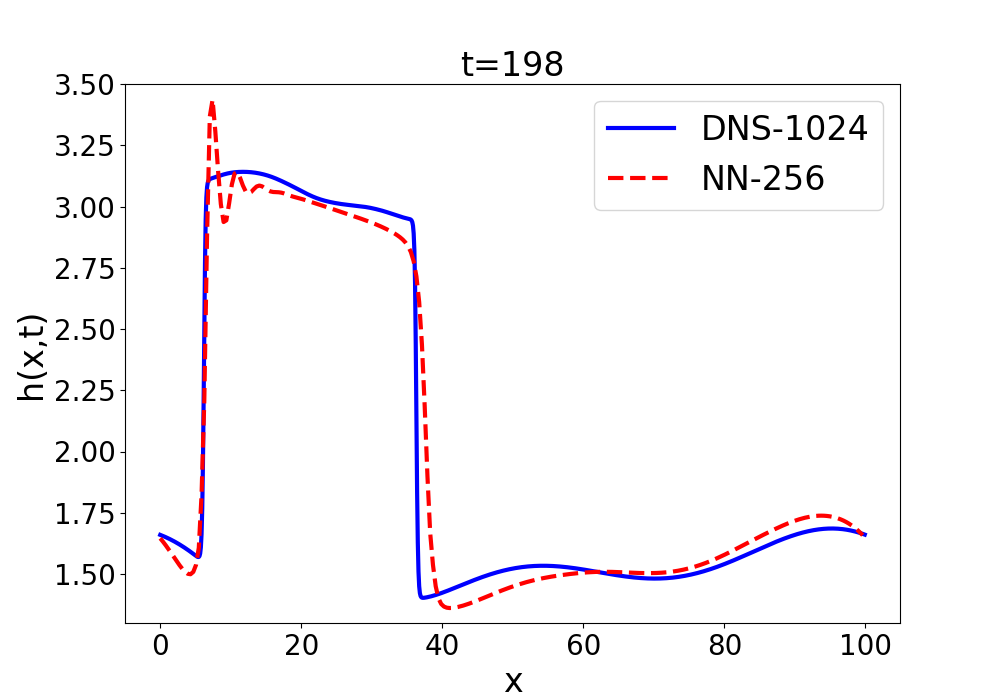}
\includegraphics[width=0.45\textwidth]{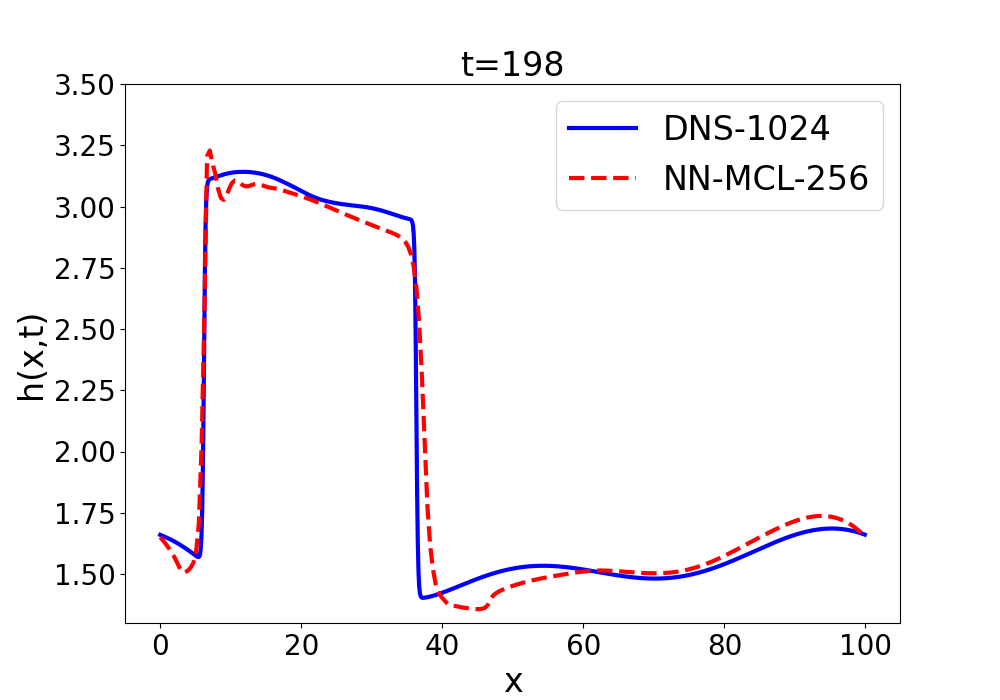}
}
\centerline{
\includegraphics[width=0.45\textwidth]{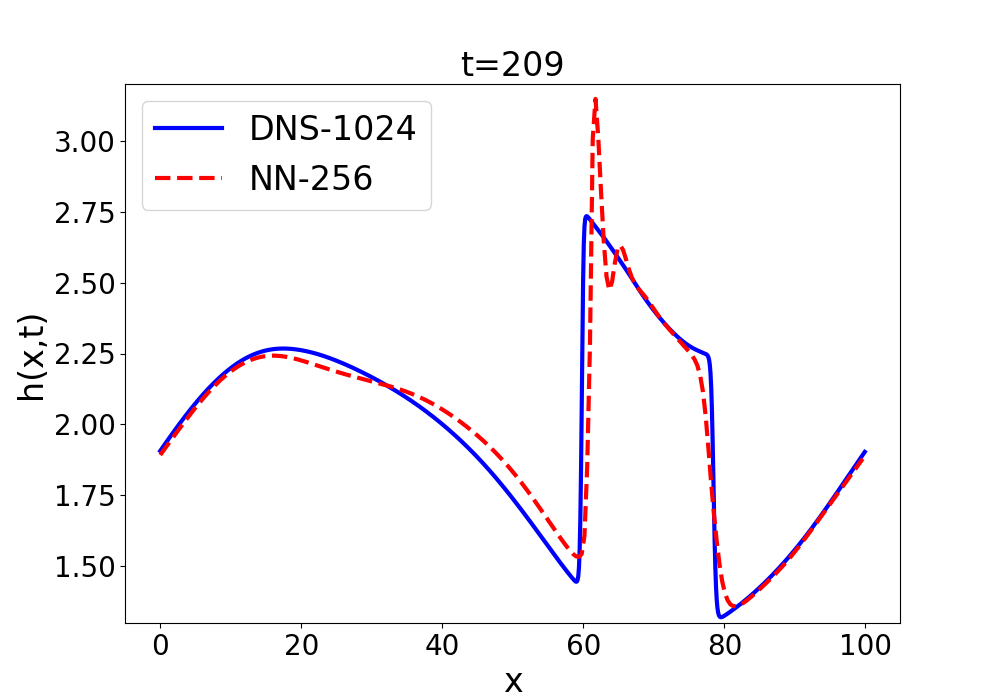}
\includegraphics[width=0.45\textwidth]{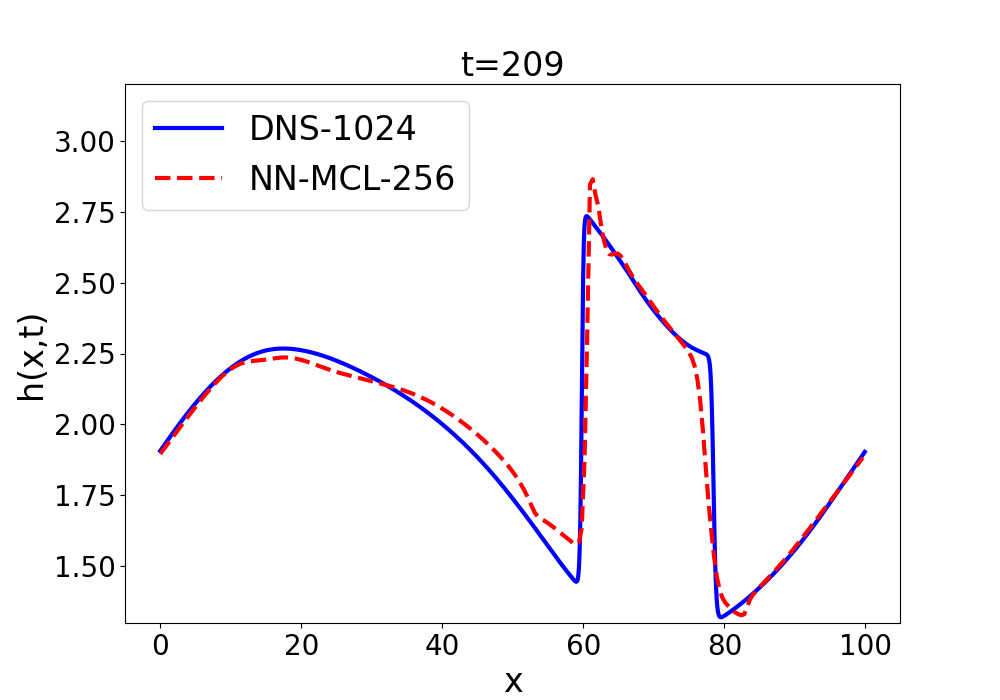}
}
\centerline{
\includegraphics[width=0.45\textwidth]{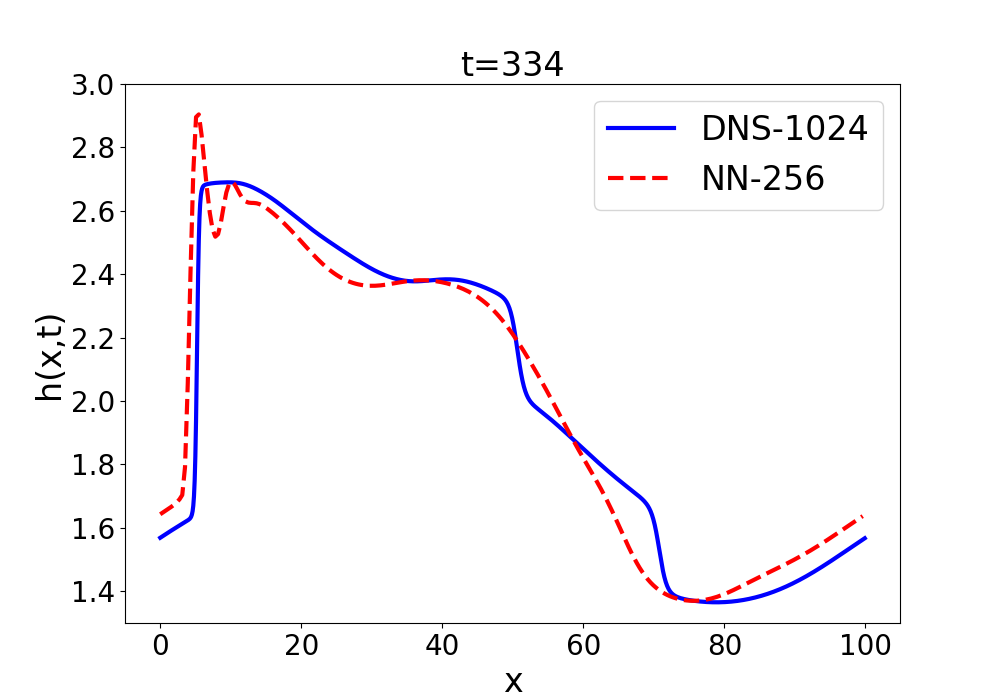}
\includegraphics[width=0.45\textwidth]{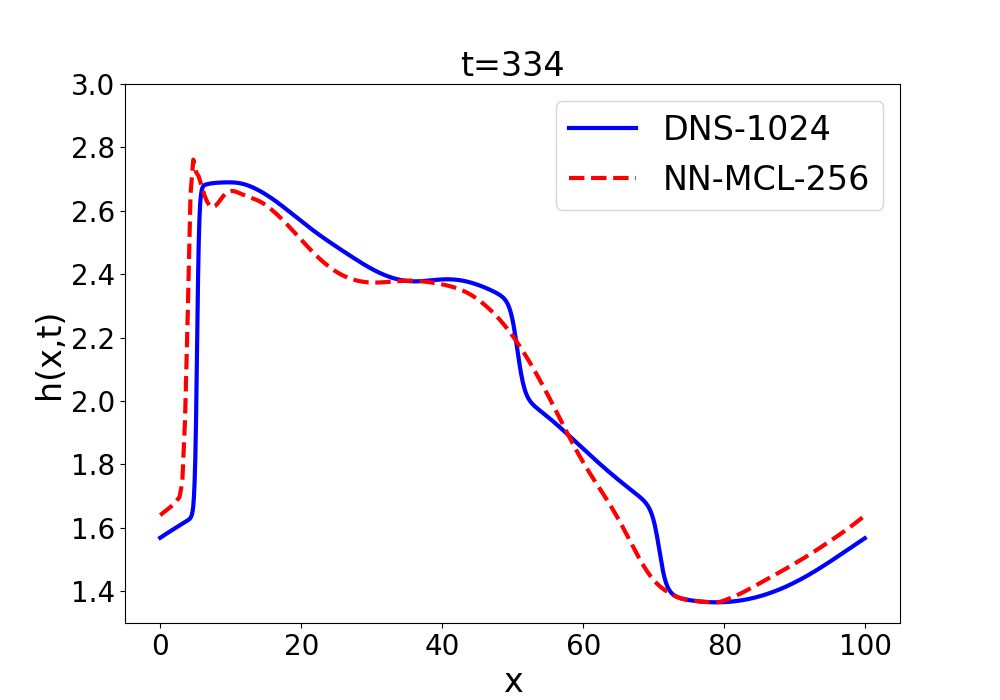}
}
\caption{Comparison between snapshots $h(x,t)$ in simulations obtained from \revision{
the reduced neural network model with 256 grid points (NN-256, red) }and \revision{the reference direct numerical simulation with 1024 grid points 
(DNS-1024, blue)} in the left column) and the neural network model 
with monotonicity-constrained learning at resolution 256 
(NN-MCL-256, red) and DNS-1024 (blue) in the right column, at times $t=198$, $209$, $334$. The NN-256 model shows very good agreement with the DNS reference, successfully capturing the hydraulic jump and the nonlinear wave propagation. The monotonicity-constrained learning (MCL) strategy is essential in reducing oscillations near shocks in simulations of the \revision{neural network} reduced model.}
\label{fig:nn256_snapshots}
\end{figure}

To further assess the performance of the NN reduced model, we compare individual solutions computed from the DNS and NN-256 with the same initial condition and forcing realization. We plot snapshots for $h(x,t)$ at times $t=198$, $209$, and $334$ in Figure \ref{fig:nn256_snapshots}. 
These snapshots correspond to stages where the flow exhibits strong nonlinear behavior, including the steepening of hydraulic jumps and sharp transitions in the free surface.
Solutions of the NN reduced model exhibit strong spurious oscillations near shocks. 
\revision{This indicates that the neural network does not reproduce fluxes at discontinuities with the same accuracy as in smooth regions. 
Since it is known that neural networks perform well when approximating continuous functions \cite{Cybenko1989, Hornik1989}, it is not surprising that the quality of the NN approximation for fluxes diminishes near shocks because the fluxes are also discontinuous at that point.
In addition, filtering of the training data (Section \ref{sec:net}) can also contribute to the discrepancies of subgrid flux estimation by the NN at the shock, since the filtered dataset does not contain data with very high gradients.}
To alleviate these oscillations, we apply the MCL limiter to the nonlinear fluxes generated by the neural network, as discussed in section \ref{sec:MCL}.
The left and right parts of the Figure 
\ref{fig:nn256_snapshots} depict snapshots without and with the MCL, respectively.
We also performed simulations of the NN reduced model with $N_c=128$ (see \cite{amranmojamder2026} for details).
The overall performance of the 
NN reduced model with $N_c=128$ is similar with the results for NN-256, except that NN-128 is more diffusive near shocks and does reproduce sharp gradients as well as NN-256, as is evident from the plot of the spectrum in
Figure \ref{fig:energy_spectra}. 
The energy spectra for both $h$ and $q$ are not affected by the MCL and are therefore not shown for brevity.

\subsection{Simulations with Topography and Manning Friction.}
To illustrate the applicability of our NN model 
outside of the training regime, we also perform simulations of SWE with bottom topography and Manning friction. 
This is a challenging test since we consider changes in the geometry of the computational domain, but do not retrain the NN parametrization.

We add the terms $[0, S_{topo} + S_{fric}]^\top$ to the right-hand side of the SWE equation \eqref{swe} and perform fully resolved and coarse simulations with the NN parametrization trained in the original regime, as outlined in Section \ref{sec:net}.
The topography and friction terms are given by 
\[
S_{topo} = -g \, h \, \partial_x b(x), \qquad
S_{fric} = -\, g\, m^2 \, \frac{q\,|q|}{h^{7/3}}, 
\]
where $b(x)$ is a known bottom topography and $m$ is the Manning’s roughness coefficient. \revision{The inclusion of these source terms follows the standard formulation of shallow water flow over variable topography with friction \cite{bermudez1994}. 
We used the numerical scheme in \cite{michel2017}; the scheme is non-negativity preserving and well-balanced (see also \cite{li2017} for a discussion on this issue).
The Manning friction is a nonlinear term that damps the momentum, thus reducing the velocity of the flow. The Manning friction can also be expressed as 
$S_{fric} = - g m^2 u|u|/h$; thus, this damping effect is particularly relevant for high-velocity flows. Since Manning friction does not depend on the gradient of the solution, it does not affect the shock structure for low to medium-velocity flows. The main consequence of including the Manning friction is the reduction of the shock magnitude. Spurious oscillations near the shock might be reduced slightly with Manning friction, but are still quite pronounced and require the application of the MCL strategy.}
We consider the 
Gaussian topography
$b(x) = A_b \exp\!\left[-B\left(x - L/2 \right)^2\right]$
with the amplitude $A_b=0.3$ and steepness $B=0.1$.
The Manning's roughness coefficient is $m=0.05$; \revision{additional examples are presented in \cite{amranmojamder2026}}.
The initial condition consists of a spatially varying free surface
$\eta(x,0)$ superimposed on the bottom topography, with the discharge $q(x,0)$
chosen so that the flow is initially subcritical, i.e., 
\begin{align*}
    & \eta_0(x)=H_0+A_h \left( \sin \left(\frac{2\pi  x}{L}+\phi_h^{(1)} \right) + \sin\left(\frac{4\pi x}{L}+\phi_h^{(2)}\right) \right),\\
    &  h_0(x)=\eta_0(x)-b(x), \qquad 
    v_0(x)=V_0, 
\end{align*}
where $x\in [0,L]$ with $L=100$. The height average $H_0=2$ is constant for all initial conditions. The remaining parameters are randomly generated following the distributions
\begin{equation*}
    V_0=\textit{Unif}[1,2], 
    \hspace{1cm} A_h=\textit{Unif}[0.1,0.6],  \hspace{1cm} 
    \phi_h^{(1)},\phi_h^{(2)}=\textit{Unif}[0,2\pi]\,,
\end{equation*}
where $\textit{Unif}[a,b]$ represents the continuous uniform distribution on $[a,b]$.

\begin{figure}[H]
\centerline{    
\includegraphics[width=0.45\textwidth]{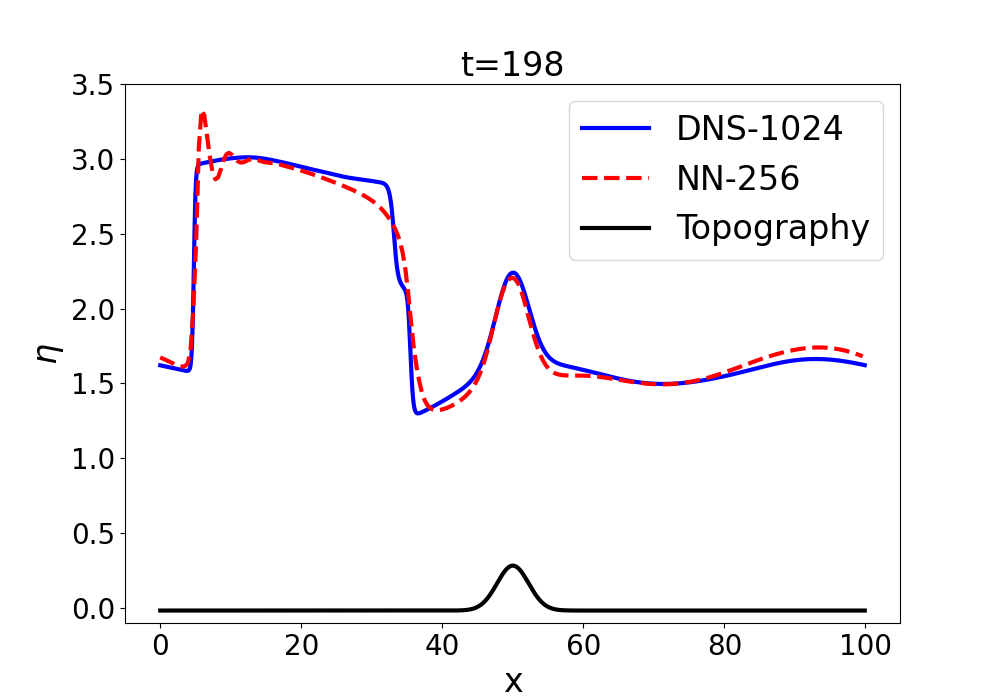}
\includegraphics[width=0.45\textwidth]{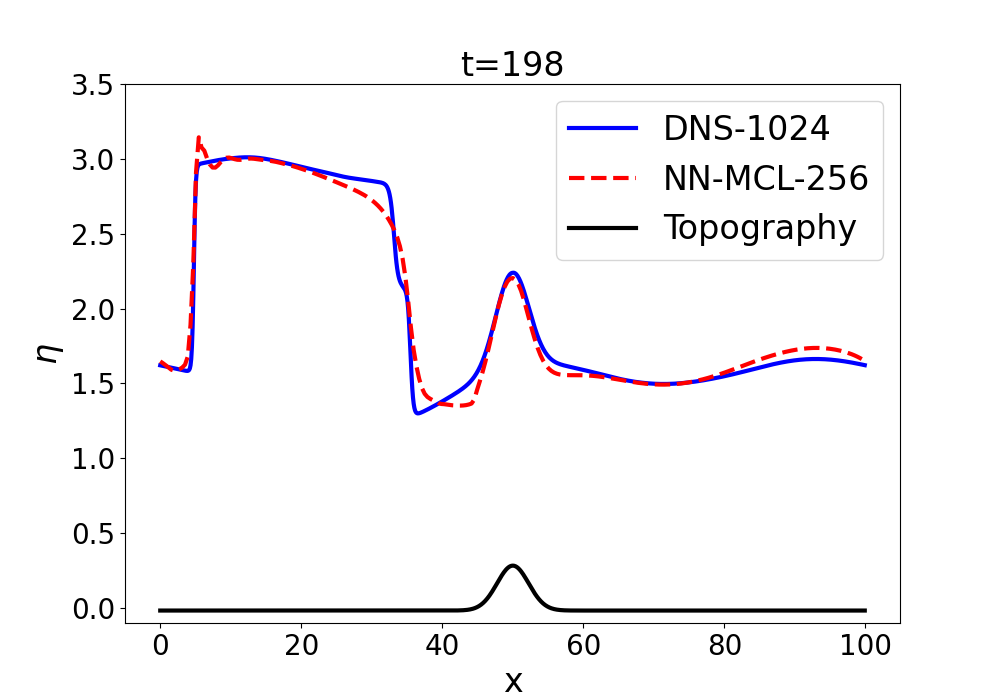}
}
\centerline{
\includegraphics[width=0.45\textwidth]{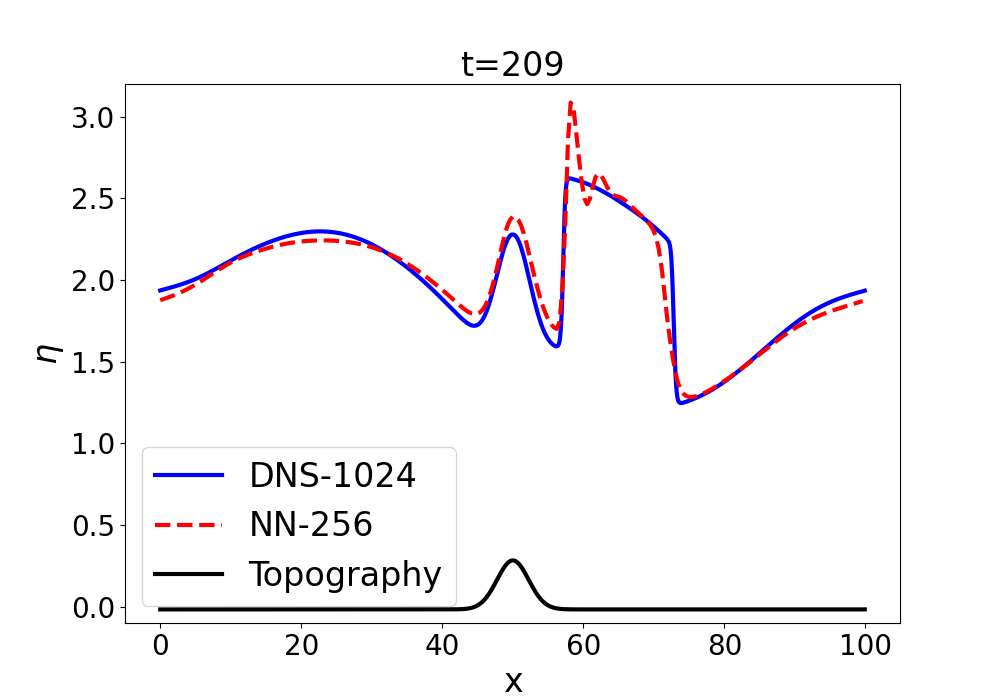}
\includegraphics[width=0.45\textwidth]{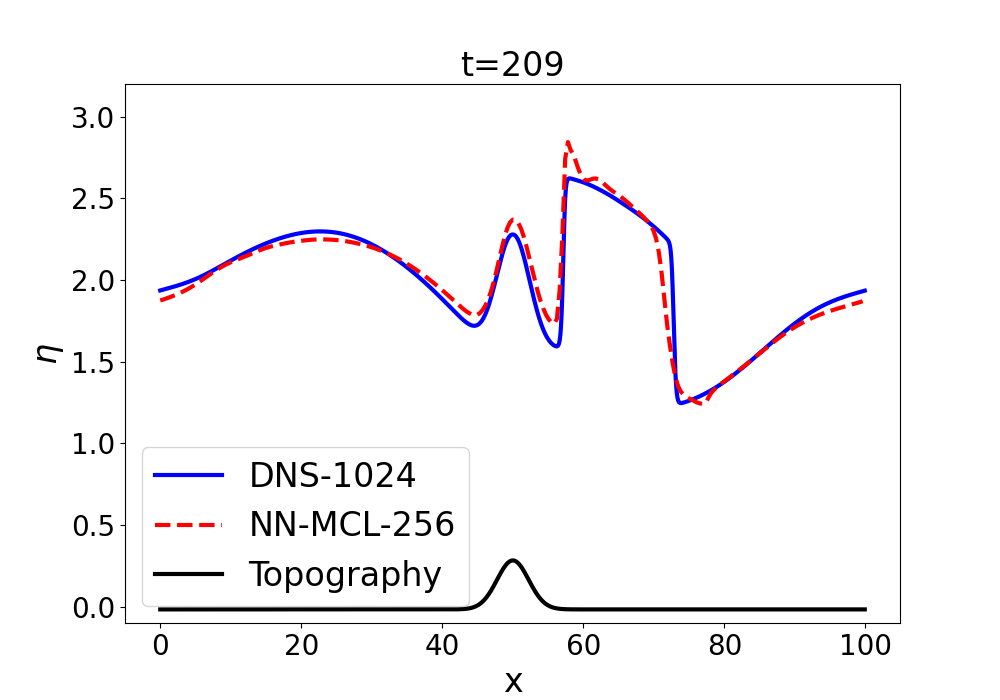}
}
\centerline{
\includegraphics[width=0.45\textwidth]{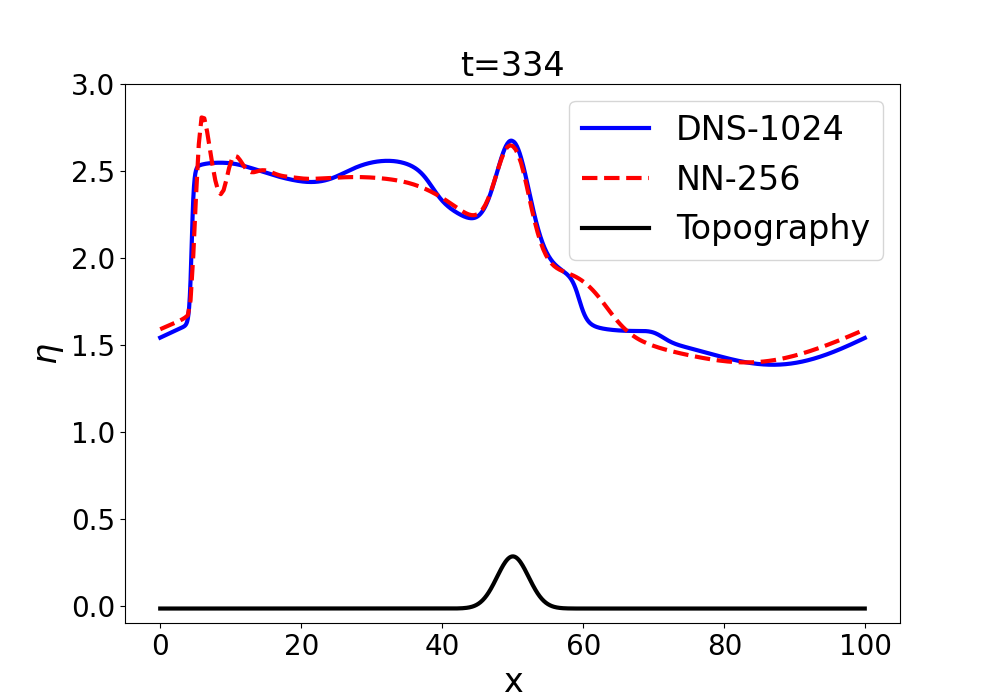}
\includegraphics[width=0.45\textwidth]{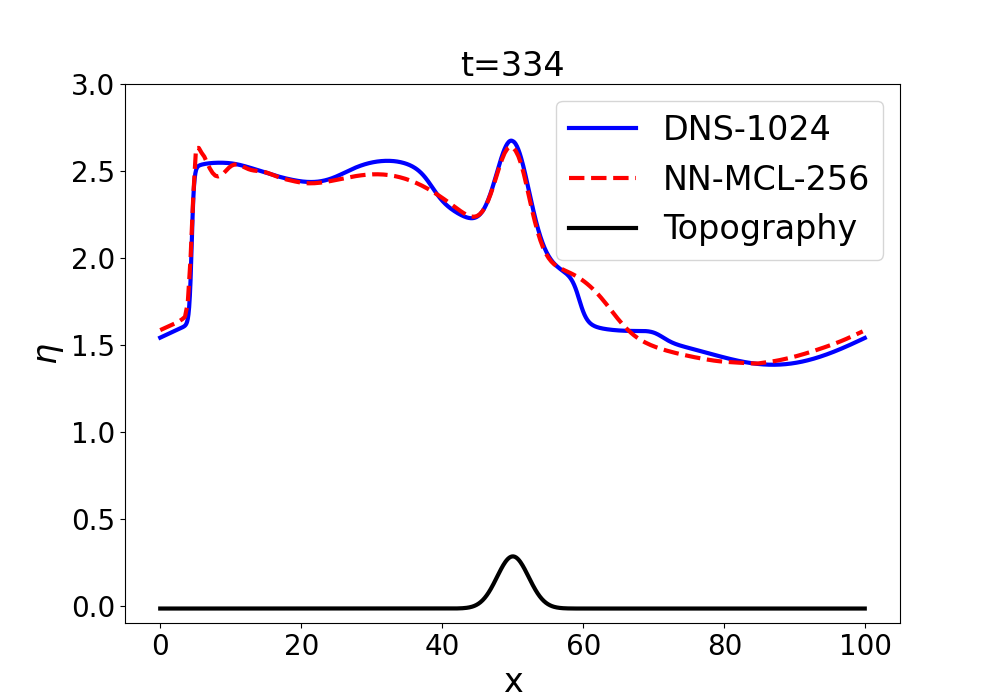}
}
  \caption{SWE simulations with topography and Manning friction. Comparison between snapshots of the free surface $\eta(x,t)$ in simulations of
 \revision{the reference direct numerical simulation with 256 grid points (NN-256, red)} and DNS-1024 (blue) (left column) and \revision{the neural network model 
with monotonicity-constrained learning at resolution 256 
(NN-MCL-256, red)} and DNS-1024 (blue) (right column) at times $t=198$, $209$, $334$.
 \revision{The monotonicity-constrained learning (MCL)} strategy is essential in reducing oscillations near shocks in simulations of the NN reduced model.
  Note that the NN parametrization was trained on data generated with $S_{topo}=S_{fric}=0$.}
  \label{fig:profiles_256}
\end{figure}

As a reference solution, we use a numerical solution computed with the LLF scheme on a fine grid with
$N_x=1024$. We consider coarse-grid simulations with $N_c=128$ and $N_c=256$.  
Figure \ref{fig:profiles_256} depicts snapshots of the free surface elevation 
$\eta(x,t) = h(x,t) + b(x)$ 
in simulations with $N_c=256$.
Coarse simulations are similar, but slightly more diffusive, as discussed previously for simulations without topography.
Snapshots for $N_c=128$ are not depicted for brevity (details can be found in \cite{amranmojamder2026}).
These results support our conclusions drawn previously. In particular, the NN reduced model reproduces the main features of
the reference solution, including the deformation of the free surface above
the topography and the sharp gradients induced by the interaction between
topography and friction.
The NN model without a limiter can exhibit small overshoots near steep fronts, whereas the NN–MCL model noticeably reduces these oscillations while preserving the
overall amplitude and phase of the solution. Moreover, similar to our previous findings,
the NN-MCL model does not significantly affect the spectra of long-term simulations.
Simulations with topography and friction show that the NN reduced model applies to a wide range of parameters outside of the training regime. We would like to point out that any \revision{machine learning} model geared towards reproducing solutions directly (e.g., PINN and its variants) \itc{would most likely fail the topography test since in those models, it is very difficult to take geometry changes into account. Changes in computational geometry result in large differences in the solution itself, and our NN coarse model is able to 
track those changes very well
(c.f. Figures \ref{fig:nn256_snapshots} and \ref{fig:profiles_256}).}

Our numerical results demonstrate that the combination of NN subgrid parametrization and convex limiting yields a stable and accurate reduced-order model even in the presence of nontrivial source terms such as topography and Manning
friction.
Moreover, since our \revision{machine learning} formalism is designed to learn fluxes in the SWE, its applicability is much broader than that of \revision{machine learning} methods that directly learn the PDE solution.

\section{Conclusions}
\label{sec:conc}
In this paper, we developed a reduced-order model for the Shallow Water Equations by using a feed-forward neural network to approximate subgrid fluxes. 
\revision{Applying machine learning techniques to learn subgrid fluxes is a novel aspect of our work. As a consequence, }
the developed approach has several advantages. In particular, the subgrid flux parametrization is local, requiring only a four-point computational stencil to compute the subgrid flux at a specific coarse location. Moreover, our \revision{machine learning} approach can be easily combined with flux limiting for hyperbolic problems. Flux limiting ensures that numerical solutions of the coarse equations remain within a physically admissible set. 
\revision{Another novel aspect of our work is using the focal loss cost function~\cite{lin2017focal} in the context of neural networks applied to partial differential equations.
Focal loss emphasizes samples that are particularly challenging for training, such as shock regions for hyperbolic problems.}

We performed several types of tests: (i) comparison of energy spectra in long-term turbulent simulations, (ii) comparison of individual solutions, (iii) simulations with a larger forcing magnitude, and (iv) simulations with bottom topography and Manning's friction.
Our results demonstrate that our 
\revision{Neural Network (NN)} parametrization improves energy exchange across scales. 
The NN reduced model is slightly more diffusive than the fully resolved simulations, which is a direct consequence of the coarse resolution of the reduced model and the need to dissipate energy at small scales.
However, the NN reduced model represents an improvement compared with the bare truncation. 
In particular, our reduced model has a better representation of the inertial range compared with the bare truncation at the same resolution. 
An averaging window $n=8$ (NN-128 \revision{reduced model} vs \revision{the fully resolved model (DNS-1024)}) is a realistic choice in most ocean applications. \revision{For instance, the ocean resolution of standard Global Circulation Models is $1^\circ$ or 100km \cite{CMIP62016}. The high-resolution eddy-permitting global ocean models have resolutions of approximately 10-25km
\cite{highresgcm2016,highresgcm2025}.}
Our results indicate that our approach can yield improvements in the development of computationally efficient coarse models of ocean dynamics.

Test (ii) evaluated the performance of our reduced model by comparing individual solutions at full and coarse resolutions. We ensured that the same forcing realization was used in both cases. Our findings show that the NN reduced model is quite accurate and reproduces individual solutions very well. Some spurious oscillations may develop near shocks, but the monolithic convex limiting significantly reduces these oscillations while not affecting the energy spectra.

Tests (iii) and (iv) demonstrate that our reduced model generalizes well to parameters outside of the training regime. These tests were performed using an NN trained in a regime with a fixed forcing magnitude, a flat bottom, and no friction.
In test (iii), the NN coarse model responds well to changes in the forcing magnitude and tracks changes in the discharge energy. 
Test (iv) is particularly challenging since it involves a different geometry of the computational domain.
Our NN reduced model reproduces the energy spectra and individual solutions in test (iv) with the same accuracy as in the training regime. This demonstrates the versatility of our approach, which is a direct consequence of the fact that we use NNs to model subgrid fluxes and not the solution itself.

Overall, we demonstrated the applicability and versatility of using NNs to model subgrid fluxes in computational fluid dynamics.
Our modeling approach preserves the structure of the equations, such as flux formulation. This leads to several advantages compared to learning the solution directly. First, our method is computationally efficient, since NN subgrid fluxes at different spatial nodes can be computed in parallel. While this is not significant for the 1D case, this can become important for 2D layered simulations. Second, our method can be easily combined with traditional flux limiting schemes to ensure admissibility of solutions, thus using the powerful existing numerical analysis machinery. Third, since we train an NN to reproduce subgrid fluxes and not solutions, our NN reduced equation applies without retraining to a wide range of parameters outside of the training regimes.

\revision{Our approach can be extended to 2D multi-layered shallow-water simulations in a relatively straightforward manner. In particular, neural networks modeling fluxes in two different directions can be trained independently for each layer, and the overall training dataset can be split into several parts for training each individual network.
The main challenges for extending our work to 2D are (i) training and testing data generation is considerably more computationally expensive, (ii) data storage can also be an issue, especially for layered shallow-water equations, (iii) the implementation issue, since the MCL strategy in 2D is more complex. However, we do not foresee any conceptual issues when extending our approach to 2D shallow water problems. 

Most climate change ocean applications utilize primitive equations, where the vertical structure becomes more important. Our method does not model the vertical transport or mixing (e.g., the diapycnal parametrization)
in the primitive equations.
We believe that after discretizing primitive equations in the vertical direction, our approach can be used in each horizontal layer; however, this requires a separate investigation, since primitive equations are often discretized in isopycnal coordinates (density-based layers).}

Our results raise several important questions for future research. One direction is to explore whether additional physical properties, such as entropy inequalities, can be incorporated into the model. Also, flux limiter constraints can be potentially included in the cost function as well. 
Since our method is suitable for conservation laws, it could be extended to other systems.
One system of particular interest is multi-layer shallow-water dynamics in 2D, which is closely related to primitive equations commonly used to simulate ocean dynamics. 
\revision{In this application, it is particularly interesting to investigate the effect of the Coriolis force, which is absent in 1D.} Another interesting case is the shallow water equations with 
wetting and drying.

\section*{Acknowledgement} 
The authors would like to thank Dr. D. Kuzmin for helpful discussions about the MCL strategy.


\end{document}